\documentclass[multphys,vecphys]{svmult}

\usepackage{makeidx}         
\usepackage{graphicx}        
\usepackage{multicol}        
\usepackage[bottom]{footmisc}

\makeindex             

\def\apj{{ApJ}}                 
\def\apjl{{ApJ}}                
\def\apjs{{ApJS}}               
\def\aap{{A\&A}}                
\def\mnras{{MNRAS}}             
\def\prd{{Phys.~Rev.~D}}        

\def\physrep{{Phys.~Rep.}}   

\newcommand{\mpc}{\rm {h^{-1}Mpc }}

\newcommand{\xiav}{\bar{\xi}}

\newcommand{\avg}[1]{\langle{#1}\rangle}

\newcommand{\ltsima}{$\; \buildrel < \over \sim \;$}
\newcommand{\lsim}{\lower.5ex\hbox{\ltsima}}
\newcommand{\gtsima}{$\; \buildrel > \over \sim \;$}
\newcommand{\gsim}{\lower.5ex\hbox{\gtsima}}

\newcommand{\citep}[1]{\cite{#1}}
 
\def\gtrsim{\mathrel{\hbox{\rlap{\hbox{\lower4pt\hbox{$\sim$}}}\hbox{$>$}}}}

\def\lesssim{\mathrel{\hbox{\rlap{\hbox{\lower4pt\hbox{$\sim$}}}\hbox{$<$}}}}

\newcommand{\tw}{\tilde{\xi}}

\begin{document}

\title*{Introduction to Higher Order Spatial Statistics in Cosmology}
\author{Istv\'an Szapudi\inst{1}
}
\institute{Institute for Astronomy, University of Hawaii, \\
2680 Woodlawn Dr, Honolulu, HI 96822\\
\texttt{szapudi@ifa.hawaii.edu}
}
%
%
\maketitle


%

\section{Introduction}
\label{sec:1}

Higher order spatial statistics characterize
non-Gaussian aspects of random fields,
which are ubiquitous in cosmology: from the Cosmic Microwave Background (CMB)
to the large scale structure (LSS) of the universe. These
random fields are rich in their properties; they can be continuous
or discrete,; can have one through three, or even more dimensions;
their degree of non-Gaussianity ranges from tiny to significant.
Yet, there are several techniques and ideas, which are applicable
to virtually all cosmological random fields, be it Lyman-$\alpha$ forests,
LSS, or CMB.

In this lecture notes I concentrate on the classic and widely
applicable characterization
of higher order statistics by joint moments, a.k.a. higher order
correlation functions, and directly related statistics. These
statistics are very powerful, although have a perturbative nature
to them in that they constitute an infinite (formal) expansion.
Clearly, when only the first $N$ terms of this expansion are extracted
from data, interesting information might remain in even higher order terms. 
This is why a host of alternative statistics (void probabilities,
wavelets, Minkowski-functionals, minimal spanning trees, phase
correlations, etc. just to name a few) are introduced and used
extensively in the literature, in complementary fashion
to $N$-point functions.More on these alternatives appear in other lecture
notes of this
volume, and if your appetite is whet you should 
read e.g., \cite{MartinezSaar2002} for more information. 
The present lecture notes serve as an informal
introduction to the subject, a starting point for 
further studies, rather than a full-blown review.

Higher order statistics are complicated due to several
factors. Most of them arise from the large configuration space
of parameters which results in a ``combinatorial explosion''
of terms. For instance, the first non-trivial $3$-point correlation
function can in principle depend on $9$ coordinates.  Taking
into account applicable symmetries still leaves
$3$ parameters in real space, and $6$ parameters in redshift
space ($5$ if one uses the distant observer approximation). 
This dependence on a large number of parameters renders
higher order statistics quite cumbersome, their measurement,
interpretation and visualization surprisingly complex and
CPU intensive. For the same reason, theoretical prediction of higher
order statistics correspondingly involved,
non-linearities, redshift distortions or projection
effects, and bias affect them in subtle and non-trivial ways.
In addition, higher order statistics are more
sensitive to geometry and systematics then their low order
cousin, the two-point correlation function. This means
that a comparatively larger and cleaner sample is needed for effective
studies. Even with the presently available large surveys, 
such as SDSS and 2dF, the
overlap between well understood theory and reliable measurement
is in fact disquietingly small. 

Despite the above difficulties, the study of higher order statistics
is a rewarding one, both intellectually and scientifically. 
The simple idea that small initial
fluctuations grew by gravitational amplification
is enough to predict higher order correlation functions, at least
on large scales, to impressive accuracy. Analytical predictions
and simulations contrasted with data have already provided a strong
support for Gaussian initial conditions, and show that our basic
picture of structure formation is correct.

On the other hand, the two-point
correlation function alone gives a quite simplistic view of LSS, which
has been visually demonstrated first by 
Alex Szalay,  through randomizing the
phases of a highly non-Gaussian simulation. 
This procedure erases a large portion of (altough not all)
higher order correlations, while keeping
the two-point statistics (c.f. Fig. 1. in Peter Coles' lecture notes
in the same volume). It is striking how
most structure we see in the simulation disappears when
higher order information is erased, despite that the two-point
function (or power spectrum) is the same. 

In addition, the two-point correlation function
is degenerate with respect to the bias 
(the difference between the statistics of the observed
``light'' from galaxies and the underlying dark matter), 
and the amplitude of primordial fluctuations.  
Most methods which extract information
from the two-point correlation function or power spectrum alone,
draw on assumptions about the bias to avoid this degeneracy problem,
and/or combine with CMB measurements.
Higher order statistics not only yields a more accurate picture
of the distribution, but resolves the degeneracy of the two-point
function by constraining the bias.

These notes focus on phenomenological aspects of higher order
correlation functions and related statistics. In particular,
I emphasize the relationship of these statistics with symmetries.
Some aspects of this are quite new, and it is the only way I know how to ease 
the pain of the above mentioned ``combinatorial explosion'', at
least partially. 
In the next sections I review the most important theoretical 
and statistical information, in particular I
present the definitions of the statistics, estimators, how
errors are calculated, algorithms to realize the estimators,
and some points on bias, redshift distortions. I develop symmetry
considerations in slightly more detail throughout.
For completeness, I
include the foundations of perturbation theory; a
detailed recent review is available by \cite{BernardeauEtal2002}
which is highly recommended. Finally, I illustrate most of the
above through an example by working out the theory of conditional cumulants. 
Finally, while LSS is emphasized in examples, most of the theory
is directly applicable to CMB, or any other random field as well.


\section{Basic Definitions}



The learning curve of higher order statistics is initially
steep, partly due to the large number of new concepts and definitions
which need to be absorbed. Here I attempt to collect
some of the most basic definitions which will be used later.

{\em A Spatial Random Field} A is field which has random values
at all spatial points. Its corresponding measure is a functional,
${\cal P}(A) {\cal D}A $, the ``probability'' of a realization. 
This is analogous to a probability density. Averages can be
calculated via functional integrals, which
although not always well defined mathematically, are very pictorial.

{\em Ensemble Average} We denote with $\avg{A}$ the ensemble
average of some quantity $A$. The averaging corresponds to a 
functional integral over the density measure. Physically 
this means simply the average of the quantity
over infinite (or many) independent realizations.

{\em Ergodicity} In the study of random fields it is often
assumed that ensemble averages can be replaced with spatial averages.
Physically this means that distant parts of the random field are
uncorrelated, and thus a large realization of the random field (a
``representative volume'', hopefully the chunk of our universe 
which is observable) can be used for measurements
of statistical quantities.

{\em Joint Moments} The most basic set of statistics we can
consider to characterize a random field $T$ are the joint moments
of the field taken at $N$ distinct points $x_1,\ldots x_N$,
\begin{equation} 
   F^{(N)}(x_1,\ldots,x_N) = \avg{T(x_1),\ldots, T{x_N}}
\end{equation} 
We often work with fluctuation fields which have been normalized with the
average: $\delta  = \frac{T}{\avg{T}}-1$. Note that the
spatial dependence on coordinate $x_i$ is often abbreviated
as $\delta_i$. No assumption is made on the dimensionality
of the random field. One (Lyman alpha forest), two
(CMB, LSS projected catalogs), or three (LSS redshift surveys) 
dimensional fields are typical in cosmology.

{\em Connected Moments} Connected moments are denoted
with $\avg{}_c$ and are defined recursively with the following
formula:
\begin{equation}
   \avg{\delta_1,\ldots,\delta_N}_c = \avg{\delta_1,\ldots,\delta_N} -
    \sum_{P}\avg{\delta_1\ldots...\delta_i}_c\ldots
     \avg{\delta_j\ldots\delta_k}_c\ldots,
\end{equation} 
where $P$ denotes symbolically a summation of all possible partitions. 
In another words, connected moment of all possible partitions have to be 
subtracted from the full (or disconnected) joint moment. What is
left is the connected moment.

{\em $N$-point correlation functions} are defined as the
connected joint moments of the normalized fluctuation field
$\delta$
\begin{equation}
   \xi^{(N)}(1,\ldots,N)=\avg{\delta_1,\ldots,\delta_N}_c,
\end{equation} 
where we introduce yet another short hand for denoting
$\xi^N(x_1,\ldots,x_N)$. The three-point correlation function
is often denoted with $\zeta$ instead of $\xi^{(3)}$, and the two-point
correlation function with $\xi\equiv\xi^{(2)}$.
Most statistics we are dealing with in this review are special
cases of $N$-point correlation functions.

{\em Gaussian} 
A random field is called Gaussian if its first and second moments
fully determine it statistically. We will often use the field
$\delta$, which denotes a random field whose average is 0. Then
a Gaussian field is fully determined by its two-point correlation
function $\xi=\avg{\delta_1 \delta_2}$. Gaussian random fields
have trivial higher order moments in that all their connected
moments vanish for $N \ge 3$ (Wick-theorem).

{\em Non-Gaussian} 
A non-Gaussian field has at least one non-vanishing higher
than second connected moment. Note that the definition of
non-Gaussian is highly general (like ``non-Elephant'') everything
what is not Gaussian is non-Gaussian. Therein lies one of the
highest difficulty of the subject.

{\em Cumulants} correspond to the simplest special case of $N$-point functions:
the joint moment of a degenerate configuration, where all the
field values are taken at the same place. The usual definition
in cosmology is
\begin{equation}
   S_N=\frac{\avg{\delta^N}_c}{\avg{\delta^2}^{N-1}},
\end{equation} 
where the normalization with the average correlation function
$\bar\xi=\avg{\delta^2}$ is  motivated by the dynamics 
of perturbation theory. Other normalizations might be more
suitable, e.g., for CMB. Cumulants 
depend only on one scale: the smoothing kernel of $\delta$ which
implicit in the above equation. This fact accounts for most of
their popularity, and indeed they are the only statistics (along with cumulant
correlators), which have been measured as high order as $N=10$
for galaxies. An alternative definition is $Q_N = S_N/N^{N-2}$,
which corresponds to the division with the number of possible
trees graphs. This definition typically ensures that for galaxies
and dark matter all $Q_N$'s are of order unity.

{\em Cumulant Correlators} correspond to the next simplest 
special case of $N+M$-point functions:
the joint moment of a doubly degenerate configuration, where the
$N$ and $M$ field values are taken at two places, respectively. 
The usual definition in cosmology is
\begin{equation}
   Q_{N,M}=\frac{\avg{\delta_1^N\delta_2^M}_c}
   {\avg{\delta^2}^{N+N-1} \avg{\delta_1\delta_2} N^{N-2}M^{M-2}}.
\end{equation} 
Other generalizations, e.g. with triple degeneracy, are obvious,
but not clear whether fruitful. In the above, it is assumed that
the smoothing kernel is the same for both points, which is usual,
although not necessary. When the normalization
of the possible number of trees is not done, the quantity
is often denoted with $C_{N,M}$.

{\em Conditional Cumulants} 
are defined as the joint connected
moment of one unsmoothed and $N-1$ smoothed density fluctuation
fields. They can be realized by integrals of the $N$-point
correlation function through $N-1$ spherical top-hat windows,
\begin{equation}
U_N(R_1, \ldots, R_{N-1})=
  \int \xi_N (x_1, \ldots, x_{N-1},0)
       \prod_{i=1}^{N-1} d^3 x_i \frac{W_{R_i}(x_i)}{V_i}
  \label{eq:cc}
\end{equation}
where $V_i = 4\pi/3 R_i^3$ is the volume of the window function
$W_{R_i}$. In the most general case, each top hat window
might have a different radius. The conditional cumulants
are  analogous to ordinary cumulants, but subtly
different: they have an advantageous smoothing kernel which enables 
their measurement with an edge corrected estimator (more on this later).

{\em Power Spectrum} Since space is assumed to be {\em statistically} 
invariant under translations, it is convenient to to introduce
the Fourier transform of the $D$-dimensional field 
\begin{equation}
   \delta(x) = \int \frac{d^Dk}{(2 \pi)^D} \delta_k e^{ikx}.
\end{equation} 
The translation invariance causes the independent $k$-modes
to be uncorrelated, which renders the correlation function
of the Fourier modes ``diagonal''
\begin{equation}
   \avg{\delta_{k_1}\delta_{k_2}} \equiv \delta_D(k_1+k_2)(2 \pi)^D P(k),
\end{equation} 
which is the definition of the power spectrum. ($\delta_D$ is the 
Dirac-$\delta$ function.) It can be shown that the power spectrum
is the Fourier transform of the two-point correlation function $\xi$.
Note that the $(2 \pi)^D$ normalization is customary, but it
is not followed by everybody.

{\em Poly-Spectra} Similarly to the power spectrum, the joint
connected moments of $N$-Fourier modes are called a poly-spectra
($N-1$-spectrum) modulo a Dirac-$\delta$ function.
The most important special case is the bispectrum, the Fourier
transform of the three-point correlation function $\zeta\equiv\xi^{(3)}$.
The next order is called tri-spectrum.

{\em A Discrete Random Field} corresponds to all possible arrangements
of discrete objects (whose number can also vary) with the corresponding
measure. A generalization over the continuous random field which is
needed to describe the statistics of the distribution objects in space,
such as galaxies.

{\em Poisson Sampling} Often it is a good approximation to derive
a discrete random field from a continuous one via a simple assumption:
take an infinitesimally small volume, such that the probability of
having more than one object in the volume is negligible, and simply
assume that the probability of having one object is proportional to
the value of the field within the small volume. This is called
an infinitesimal Poisson process.

{\em Discreteness Effects} correspond to the difference between the
statistics of an observed, discrete random field  and an assumed 
underlying, continuous random field. Also known as shot noise, or
Poisson noise. It can be especially simply calculated under the
assumption of infinitesimal Poisson sampling.


\section{Estimators}

The main idea underlying the construction of estimators
for spatial statistics is the ergodic theorem. If we
replace the ensemble averages with spatial averages,
we obtain estimators for the above defined quantities.

To put the above simple idea into practice, one has
to deal with some subtleties as well.
I list a few interesting ones: discreteness, counts in cells,
edge correction, and optimality.

\subsection{Discreteness}

Imagine a galaxy catalog with average count $\bar N$ in
a given cell of size $R$, our chosen smoothing length.
We can estimate approximately the density at each cell
as $\tilde\delta_i\simeq\frac{N_i}{\bar N}-1$.
The above simple idea suggest to estimate the $K$-th order
cumulant with (for a quantity $A$ we denote
its estimator $\tilde A$ throughout). 
\begin{equation}
   \tilde S_K \propto \frac{1}{N_{tot}}\sum {\tilde\delta_i}^K|_c.
\end{equation}
A little more thought, however, reveals that, while our estimator for
density field is unbiased, our estimator for the cumulant is not.
While $\avg{N} = \bar N \avg{1+\delta} = \bar N$, $\avg{N^K} 
\ne {\bar N}^K \avg{(1+\delta)^K}$. The reason for this is
the ``self contribution'' to correlations, a typical discreteness
effect.

To see this, let's follow \cite{Peebles1980}, and
imagine for $K=2$ that our cell of size $R$ is divided
into $T$ infinitesimally small cells, each of them small enough
that the probability of having more than one galaxy in it would
be negligible. Let's call the number of galaxies in each tiny
cells $\mu_i$. Since $\mu_i = 0,1$, $\mu_i^K = \mu_i$ for any $K$.
This means that all moments $\avg{\mu_i^K}=\avg{\mu_i}= \bar N/T$.
The total number of objects in our original cell is $\sum \mu_i$.
Now it is easy to see that
\begin{equation}
   \avg{N^2} = (\sum\mu_i)^2 = \sum_i\avg{\mu_i}+\sum_{i \ne j}
   \avg{\mu_i}\avg{\mu_j}(1+\xi_{ij}),
\end{equation}
With $T \rightarrow\infty$ the above expression yields the
final results
\begin{eqnarray}
   \avg{N^2} =& \bar N + {\bar N}^2(1+\xiav) \cr
             =&  \bar N + {\bar N}^2\avg{(1+\delta)^2}.
\end{eqnarray}
The first term corresponds to the Poisson-noise bias of our estimator.
Note that we already simplified the
case assuming that $\bar N$ and all the normalization required
to obtain $S_K$ from the above is known a priori.

Higher orders can be calculated either with the above suggestive,
but tedious method, or with generating functions one can prove     
the following simple rule \cite{SzapudiSzalay1993}
\begin{equation}
   \avg{(N)_K} = {\bar N}^K\avg{(1+\delta)^K},
\end{equation}
where $\avg{(N)_K} = \avg{N(N-1)\cdots(N-K+1)}$ are the factorial moments
of $N$ (the quantity inside the average is called falling factorial). 
In other words, if you replace regular moments with factorial
moments, discreteness effects magically disappear. With this
simple rule, and taking into account that discreteness 
can only affect overlap, you can construct estimators free 
from discreteness biases for any of our defined estimators.
E.g.
\begin{eqnarray}
   \avg{(N)_2} =& {\bar N}^2(1+\xiav)\cr
   \avg{(N)_3} =& {\bar N}^3\avg{(1+3\delta+3\delta^2+\delta^3} \cr
               =& {\bar N}^3(1+3\xiav+3S_3\xiav^2)
\end{eqnarray}
and all you have to do is express algebraically $S_3$ as a function
of the factorial moments. It is a good exercise, for solutions
see \cite{SzapudiEtal2000}.

Note that both intuition and 
halo models \cite{CooraySheth2002} suggest that Poisson
sampling might not be a good approximation on very small scales
for galaxy surveys. When in doubt (i.e. no well defined and
trusted model for the discreteness exist), try to use estimators
without self-overlap, since discreteness manifests itself through
self-correlations only.

\subsection{Counts in Cells}

We have shown in the previous subsection how to construct
estimators for cumulants, cumulant correlators, etc. even in
the discrete case. For cumulants, however, a more efficient
method exists based on counts in cells.

According to the previous subsection, if we can 
estimate factorial moments $(N)_K$, it is a simple matter
of calculation to obtain cumulants. While one could
directly estimate the factorial moments from data, the
usual procedure is to estimate first the probability $P_N$
that a cell of a given volume (or area) contains $N$ objects.
Then the factorial moments can be calculated as
\begin{equation}
   \avg{(N)_K} = \sum_{N \ge 0} P_N (N)_K.
\end{equation}
The advantage of this technique is that there exist
many fast and efficient algorithms to estimate $P_N$ from
data, and once the counts in cells probabilities are estimated,
arbitrary high order cumulants can be obtained in a straightforward
manner.

\subsection{Edge Correction and Heuristic Weights}

If signal and noise were uniform, uniform weighting of objects
would be optimal as required by symmetry.
On scales approaching the characteristic size of a survey, uneven 
(or more generally suboptimal) weighting
of objects near the edges of the survey will often dominate the errors.
The optimal weight has contributions from survey 
geometry and correlations (signal
and noise), and it is not even diagonal.
These two effects are only separate
conceptually, but they are often coupled in non-trivial ways.

Edge correction with heuristic weights fully accounts for the
effect of survey geometry (including uneven sampling, etc),
however, it does not include a prior on the correlations.
For instance,
the heuristic weights used in the SpICE method \cite{SzapudiEtal2001}
are exact in the noise dominated regime, however, they are only
approximate when the signal is important.

A class of edge corrected estimators 
\cite{SzapudiSzalay1998,SzapudiSzalay2000} can be written symbolically as 
\begin{equation}
  \tilde \xi^{N} = \sum_{i_1,\ldots,i_N}w_{i_1,\ldots,i_N}
  (\delta_{i_1}\ldots\delta_{i_N})_c
  \label{eq:npestimator}
\end{equation}
which is formally the same as the quantity we want to measure,
except we replace $\avg{}$ with an average over the sample
itself according to the ergodic principle.
The heuristic weights are such that their total sum is $1$, and
usually correspond to an inverse variance.
This would be exact if the bins
were uncorrelated, which is usually not the case.  
It can be a good approximation power spectra
as shown by \cite{FeldmanEtal1994} (c.f. lecture notes
of Andrew Hamilton in this volume).

Note that above estimator introduce additional subtleties due to
taking of connected moments, and due to the non-linear nature
of the estimator. The non-linearity is introduced by calculating
the fluctuations via a division with the average density. If we
estimate the average density from the same sample, the bias
due to this effect is termed the ``integral constraint''.

In practice, the above estimator is often replaced with a 
Monte Carlo version (\cite{LandySzalay1993} for the two-point
correlation function, and \cite{SzapudiSzalay1998,SzapudiSzalay2000} for the
$N$-point functions). Let $D$, and $R$ denote the data set,
and a random set, respectively. The latter is a random
distribution over the geometry and sampling variations in the survey.
If we define symbolically an
estimator $D^pR^q$, with $p+q = N$ for a
function $\Phi$ symmetric in its arguments
  \begin{equation}
        D^pR^q = \sum\Phi(x_1,\ldots,x_p,y_1,\ldots,y_q),
  \end{equation}
with $x_i \ne x_j \in D, y_i \ne y_j \in R$. As an example, the
two point correlation function corresponds to
 $\Phi(x,y) = [x,y \in D, r \le d(x,y) \le r+dr]$,
where $d(x,y)$ is the distance between the two points, and
$[condition]$ equals $1$ when $condition$ holds, $0$ otherwise.

The estimator for the
$N$-point correlation functions is written with
the above symbolic notation as $(\hat D-\hat R)^N$, or
more precisely,
  \begin{equation}
        \tw_N = \frac{1}{S}  \sum_i
        {N \choose i}(-)^{N-i} (\frac{D}{\lambda})^i (\frac{R}{\rho})^{N-i},
   \label{eq:ssnpt}
  \end{equation}
where $S = \int\Phi\mu_{N}$, a simple phase space integral,
and $\lambda, \rho$ are the densities of data and random sets.
(for details see \cite{SzapudiSzalay2000}). This is the Monte
Carlo version of the edge corrected estimator.

It is worthwhile to note is that counts in cells cannot be properly
edge corrected, although approximate procedures exist in the
regime where the probabilities have a weak dependence on the
cell shape, or one could integrate the above estimators
numerically.

\subsection{Optimality}

For the measurement of the two-point function for a Gaussian
process we know that maximum likelihood methods are optimal
in the sense that no other method can achieve smaller variance.
This can be shown with the use of the Fisher matrix and 
the Cramer Rao inequality \cite{BondEtal1998, TegmarkEtal1998}
(see also the lecture notes of Andrew Hamilton in this volume)
Since our previous prescription for edge effects
is different (in particular it does not know about correlations), 
it is clearly suboptimal. Given that we know the optimal
method, why don't we use it for all measurements? The reason is that
there are caveats, which, especially for galaxy surveys, severely
limit the applicability (and optimality) of the optimal method.

\begin{itemize}

\item Computational issues:
The quadratic incarnation of the maximum likelihood
method for two-point correlation
function (or power spectrum) amounts to sandwiching the projection matrix
$P_l = \partial C/\partial l$, the derivative of the full (signal
plus noise) correlation matrix according to the $l-th$ parameter
$C_l$ between the data vectors weighted by
the inverse correlation matrix $C^{-1}$, yielding 
\begin{equation}
   \tilde C_l = < C^{-1} x | P_l | C^{-1} x >,
\end{equation}
where $\tilde C_l$ is the estimator for $C_l$. We can see
from this that the inverse of the correlation matrix needs to
be calculated, which typically scales as $O(N^3)$  with
the number of data items\footnote{There are methods, such as
preconditioned conjugate gradients, with which $C^{-1}x$ can be
calculated in $O(N^2)$. However, the Fisher matrix still needs
$O(N^3)$ computations}. 
For $N\simeq 1000$ this can be done
on a laptop computer (about 17 sec on mine), 
but for $N \gtrsim 10^6$ it becomes prohibitive
even for the fastest supercomputers (not to mention storage
capacity, which might also become unrealistic for a large data set).
Iterative, and brute force maximum likelihood methods, although seemingly 
more complicated, scale the same way.

State of the art data sets (``mega-pixel'' CMB maps, such as WMAP and Planck,
as well as decent LSS surveys) are way beyond present day computational
capabilities (unless some special tricks and
approximations are involved, e.g.,\cite{osh99}).

\item Caveats: the ``lossless and optimal'' property of the maximum likelihood
estimator is subject to practical limitations. Data reduction and analysis
involve many subtle choices, such as systematic corrections,
binning, pixellization, etc. In the past these turned into
differences between analyses which all used
the same optimal method. In addition,
non-iterative quadratic maximum likelihood estimators are
posteriors based on a prior, thus they are only optimal
in the limit when we have full prior knowledge of what we
want to measure.

\item The Gaussianity condition: while it is an excellent approximation for
the CMB, it breaks down for the LSS even on very large scales. In the
Bayesian sense, the Gaussian prior is not justified, thus the estimators
might become sub-optimal, and might even be biased. 
Those measurements, where variants of the maximum likelihood
are method are used for LSS, take great pain at controlling ``non-linear
contamination'', e.g. with filtering. Besides the computational
aspects, this is the main reason why many LSS analyses still use heuristic
weights.  However, some aspects of the maximum likelihood method have been
adapted to deal with non-Gaussianities \cite{Hamilton2000}, and
even for estimating three-point correlation functions
under the assumption of Gaussianity \cite{SantosEtal2003,HeavensEtal2003}
(this is not a contradiction: when the non-Gaussianity is tiny,
as in CMB, correlation matrices can be well aproximated with a  Gaussian
assumption).

\end{itemize}

\section{Errors}

The {\em} formal calculation of errors (or more generally
covariance matrices) is straightforward from
Equation~\ref{eq:npestimator}: one simply takes the square
of the equation and performs ensemble averages. In general,
for two bins represented by the weights $w^a$ and $w^b$,
the errors will scale as
\begin{equation}
   \Delta \tilde \xi^{N} = \sum_{i_1,\ldots,i_N}w^a_{i_1,\ldots,i_N}
w^b_{j_1,\ldots,j_N} 
\avg{\delta_{i_1}\ldots\delta_{i_N}\delta_{j_1}\ldots\delta_{j_N}}.
\end{equation}
In practice, complications will arise from i) the large number of
terms ii)  the complicated summation (or integral in the continuum
limit) iii) the overlaps between indices and the corresponding
discreteness terms iv) the theoretical expression
of the ensemble averages. However, we can draw the general idea
from the expression that the covariance matrix of $N$-point 
estimators depends on $2N$-point correlation functions. Next
we demonstrate this formula with the calculation of the errors
on the two-point correlation function.

To specify more our symbolic estimator, let us assume that 
the survey is divided into $K$ pixels, each of
them with fluctuations $\delta_i$, with $i$ running from $1\ldots K$.
For this configuration our estimator can be written as
\begin{equation}
   \tilde \xi = w_{12}\delta_1\delta_2.
\end{equation}
The above equation uses a ``shorthand'' Einstein convention:
$1,2$ substituting for $i_1,i_2$, and repeated indices summed,
and it is assumed that the two indices cannot overlap.
 
The ensemble average of the above estimator is clearly
$w_{12}\xi_{12}$. The continuum limit (co)variance between
bins $a$ and $b$ can be calculated
by taking the square of the above, and taking the ensemble average.
 \begin{equation}
   \avg{\delta \tilde \xi^a \delta \tilde \xi^b} = w_{12}^aw^b_{34}\left(
   \avg{\delta_1\delta_2\delta_3\delta_4}-
   \avg{\delta_1\delta_2}\avg{\delta_3\delta_4}\right).
  \end{equation}
Note that the averages in this formula are not connected moments,
which are distinguished by $\avg{}_c$.
 
The above equation yields only the continuum limit terms. To add
Poisson noise contribution to the error, note
that it arises from the possible overlaps between the indices
(indices between two pairweights {\em can} still overlap!).
In the spirit of infinitesimal Poisson models, we replace
each overlap with a $1/\lambda$ term, and express the
results in terms of connected moments. There are three
possibilities, i) no overlap (continuum limit)
\begin{equation}
   w_{12}^aw^b_{34}
    \left(\xi_{1234}+\xi_{13}\xi_{24}+\xi_{14}\xi_{23}\right),
\end{equation}
ii) one overlap (4 possibilities)
 \begin{equation}
    \frac{4}{\lambda}w_{12}^aw^b_{13} \left(\xi_{123}+\xi_{23}\right),
  \end{equation}
iii) two overlaps (2 possibilities)
 \begin{equation}
    \frac{2}{\lambda^2}w_{12}^aw^b_{12} \left(1+\xi_{12}\right),
  \end{equation}
In these equations, for the sake of the Einstein convention
we used $\xi(i,j,k,l) \rightarrow \xi_{ijkl}$. The above
substitutions (rigorously true only in the infinitesimal Poisson
sampling limit) become increasingly accurate with decreasing
cell size. For the Monte Carlo estimator of Eq.~\ref{eq:ssnpt}
a slightly more general formula is valid, where the summation
is replaced with integrals over a bin function $\Phi(1,2)$
\begin{eqnarray}
  \label{eq:twopoint}
  \avg{\delta\tw_2^a\delta\tw_2^b} &&= \frac{1}{S^2}\Bigr\lbrace\Bigr.
  \int\Phi_a(1,2)\Phi_b(3,4)
   \left[ \xi_4(1,2,3,4)+2\xi(1,3)\xi(2,4)\right] + \nonumber \\
  && \frac{4}{\lambda}\int\Phi_a(1,2)\Phi_b(1,3)
   \left[ \xi(2,3)+\xi_3(1,2,3)\right] + \nonumber \\
   &&\frac{2}{\lambda^2}\int\Phi_a(1,2)\Phi_b(1,2)
   \left[ 1+\xi(1,2)\right] \Bigl.\Bigr\rbrace .
\end{eqnarray}
For completeness, we present the result for the three-point 
correlation function as well
\begin{eqnarray}
  \avg{\delta\tw_3^a\delta\tw_3^b} &&= \frac{1}{S^2}\Bigl\{\Bigr. 
   \int\Phi_a(1,2,3)\Phi_b(4,5,6)\bigl[\bigr. \xi(1,2,3,4,5,6)+ \nonumber \\
  &&3\xi(1,2)\xi(3,4,5,6) +9\xi(1,4)\xi(2,3,5,6)+ \nonumber \\
  &&3\xi(4,5)\xi(1,2,3,6) +9\xi(1,5,6)\xi(2,3,4)+ \nonumber \\
  &&9\xi(1,4)\xi(2,3)\xi(5,6) +6\xi(1,4)\xi(2,5)\xi(3,6)\bigl.\bigr]
  \nonumber \\
  &&\frac{9}{\lambda}\int\Phi_a(1,2,3)\Phi_b(1,4,5)
     \bigl[\bigr. \xi(1,2,3,4,5)+ \nonumber \\
  &&\xi(2,3,4,5) +2\xi(1,2)\xi(3,4,5)+ \nonumber \\
  &&2\xi(1,4)\xi(2,3,5) +\xi(2,3)\xi(1,4,5)+ \nonumber \\
  &&4\xi(2,5)\xi(1,3,4) +\xi(4,5)\xi(1,2,3)+ \nonumber \\
  &&\xi(2,3)\xi(4,5)+2\xi(2,4)\xi(3,5)\bigl.\bigr] +\nonumber \\
  &&\frac{18}{\lambda^2}\int\Phi_a(1,2,3)\Phi_b(1,2,4)
     \bigl[\bigr. \xi(1,2,3,4)+ \nonumber \\
  &&2\xi(1,3,4) +\xi(1,2)\xi(3,4)+ \nonumber \\
  &&2\xi(1,3)\xi(2,4) + \xi(3,4) \bigl.\bigr] \nonumber \\
  &&\frac{6}{\lambda^3}\int\Phi_a(1,2,3)\Phi_b(1,2,3)
     \bigl[\bigr. \xi(1,2,3)+
  3\xi(1,2) +1\bigl.\bigr] \Bigl.\Bigr\}. 
\end{eqnarray}
In the above formula, $\xi$ with $N$ variables is an $N$-point
correlation function, $\Phi_a$ is a bin function corresponding
to a triangle (it's value is one, when the triplet is in the
bin, 0 otherwise).

The above integral is fairly complicated to calculate in
practice, usually some approximations are necessary. In series
of papers \cite{SzapudiColombi1996,SzapudiEtal2000,SzapudiEtal2001}
has worked out practical approximations for up to $N=4$ for
the moments of CIC, obtained (lengthy) semi-analytical
formulae, and checked validity of those approximations
against simulations. The resulting code ({FOrtran for Cosmic
Errors, FORCE} is available publicly.
\footnote{http://www.ifa.hawaii.edu/users/szapudi/force.html} 
The errors of CIC have the special property that they contain
terms $\propto 1/C$, where $C$ is the number of cells used
for the CIC estimation (measurement errors). This is the
motivation for algorithms with have $C=\infty$.

From the above error calculations an intuitive
physical picture has emerged. We can distinguish three classes
of errors, often approximately separated in the final expressions:
finite volume errors (the term cosmic variance is used in
CMB literature), corresponding to the the fact that the
universe might have fluctuations on scales larger than the 
survey size (the smallness of such fluctuations
often termed qualitatively as ``representative volume'');
discreteness errors: arising from the fact that we are sampling
the dark matter distribution (presumably continuous) at finite
points with galaxies; edge effect errors, arising from the
uneven weights given to galaxies during the estimation.
For typical galaxy surveys, edge effects dominate on large scales, 
and discreteness
on small scales. Finite volume effects change slower, and they
might dominate in the transition region on intermediate scales.
In the above example for the two-point function, the
first term contains mainly finite volume effects (the integral
of the 4-point function), and terms with powers of $1/\lambda$
are discreteness effects. Edge effects are due to the complicated
weight summation blended with the other effects. 

These qualitative observations are valid for CMB as well with
some caveats: i) the CMB is Gaussian to good approximation,
therefore typically only two point functions need to be taken
into account ii) the CMB is continuous, i.e. there are no
discreteness effects, iii) instead, there are terms
arising from the noise of the detection with  similar
role to discreteness (noise) in LSS.
Note that for high precision cosmology applications 
constraining cosmological parameters, 
``error on the error'' (or uncertainty of the uncertainty 
as sometimes called) is as important as the size of
the error.

\section{Symmetry Considerations}
\label{sec:sym}

A class of functions subject to (Lie-group) symmetries
is best represented as an expansion over 
irreducible representations.
In Euclidean (flat) space translation
invariance is the most obvious symmetry.
The appropriate transform is the Fourier transform, and homogeneity
can be taken into account as a Dirac delta function in transform
space. The customary definition of the power spectrum and bispectrum
are with the Fourier transform of the random fluctuation field $\delta(k)$
as
\begin{eqnarray}
   \avg{\delta(k_1)\delta(k_2)} &= (2\pi)^D \delta_D(k_1+k_2)P(k_1)\cr
   \avg{\delta(k_1)\delta(k_2)\delta(k_3)}
  &= (2\pi)^D \delta_D(k_1+k_2+k_3)B(k_1,k_2,k_3),
\end{eqnarray}
where $\delta_D$ is a Dirac delta function, and $D$ is the spatial
dimension. Thus, because of homogeneity,
the two-point function becomes the Fourier transform of the
power spectrum
\begin{equation}
  \xi(x_1,x_2) = \int \frac{d^Dk}{(2\pi)^D}P(k) e^{i k(x_1-x_2)},
\end{equation}
and a similar equation is true for the three-point correlation function
\begin{equation}
  \xi(x_1,x_2,x_3) = \int\Pi_{i=1}^3\frac{d^Dk_i}{(2\pi)^D}B(k_1,k_2,k_3)
  e^{i(k_1x_1+k_2x_2+k_3x_3)}\delta_D(k_1+k_2+k_3).
\end{equation}
From these equations, the two-point correlation function is only
a function of the difference of its two arguments. If the statistical
properties of the underlying field are isotropic as well, these
equations can be further simplified. We quote the results for
two and three spatial dimensions:
\begin{eqnarray}
   \xi(r)& = \int \frac{k dk}{2\pi}P(k) J_0(kr) \,\, \rm{2D},\cr
   \xi(r)& = \int\frac{k^2 dk}{2\pi^2} P(k) j_0(kr) \,\, \rm{3D},
    \label{eq:2ptpk}
\end{eqnarray}
where $J_0$ and $j_0$ are ordinary and spherical Bessel functions,
respectively.

As a consequence of spherical symmetry the three-point
correlation function and the bispectrum,
depend only on the shape of a triangle.
\cite{Szapudi2004a} has observed that the three-point statistics
can be expressed with two unit vectors, thus zero
angular momentum bipolar expansion is suitable in 3 spatial dimensions
under SO(3) symmetry. Zero angular momentum bipolar functions are proportional
to the Legendre polynomials, thus in turn this becomes multi-pole
expansion of the bispectrum. If we parameterize the bispectrum
as depending on two vectors, and the angle between, this can
be written as
\begin{equation}
  B(k_1,k_2,\theta)= \sum_l B_l(k_1,k_2)P_l(\cos\theta)\frac{2 l +1}{4 \pi},
  \label{eq:3ptbl3d}
\end{equation}
with an entirely similar expansion for the three-point correlation
function. 
It is then simple matter to show \cite{Szapudi2004a} that
the multi-poles of the bispectrum are related to the multi-poles
of the three-point correlation function via a double Hankel transform
\begin{equation}
   \xi^3_l(r_1,r_2) = \int\frac{k_1^2}{2 \pi^2}dk_1 \frac{k_2^2}{2 \pi^2} dk_2
    (-1)^l B_l(k_1,k_2)j_l(k_1r_1)j_l(k_2r_2).
\label{eq:transform3D}
\end{equation}
$B_l$'s and $\xi_l$ are analogous to the $C_l$'s of the angular
power spectrum, they correspond to an angular power
spectrum of a shells of three-point statistics.

In two dimensions, the situation is entirely analogous:
the symmetry is just U(1) rotations on a ring, thus
the correct expansion if Fourier (actually cosine) transform
\begin{equation}
  B(k_1,k_2,\theta)= \frac{B_0(k_1,k_2)}{2}+
  \sum_{n <  0} B_n(k_1,k_2)\cos(n\theta),
  \label{eq:3ptbl2d}
\end{equation}
with analogous expansion for the three-point correlation functions.
Note that the $\sin$ modes are killed by parity.
This expansion is generally applicable for any
three-point statistics in isotropic 2-dimensional spaces.
This in turn renders the connection of the Fourier coefficients
of the three-point correlation function and bispectrum as
\begin{equation}
   \xi^3_n(r_1,r_2) = \int\frac{k_1}{2 \pi}dk_1 \frac{k_2}{2 \pi} dk_2
    (-1)^n B_n(k_1,k_2)J_n(k_1r_1)J_n(k_2r_2).
\label{eq:transform2D}
\end{equation}
This latter equation, derived for two-dimensional Euclidean
space, is also applicable to the spherical bispectrum in the
flat sky approximation. This
provides surprisingly accurate results,
even for relatively law multi-poles.
A form equivalent to the expansion of Eq.~\ref{eq:transform2D}
was proposed independently by \cite{Zheng2004} as well,
for the (redshift space) projected bispectrum; this is
yet another application of the flat sky approximation.

\section{Algorithms}

Estimation higher order statistics is a daunting
computational problem.  Naive solutions are typically
ranging from the inefficient (for two-point correlations
and counts in cells) to the impossible, hence the
application of advanced computer science is a must.

For counts in cells, the naive algorithms scales 
as $M N_C$ where is preferably $N_C \gtrsim 10^{9-12}$, the number
of cells used for the estimation.

Naive estimation estimation of $N$-point quantities
typically scales $M^N$ where $M$ is the number of
data points. For modern data sets $M \gtrsim 10^{6-10}$,
this becomes prohibitive for higher than second order
correlation functions. This is a fast developing field.
Next we present a set of algorithms to illustrate the
problems and typical solutions.

\subsection{$N$-point Correlation Functions}

\subsubsection{Hierarchical Algorithms}
\begin{figure} [htb]
\centering
\includegraphics[width=6cm]{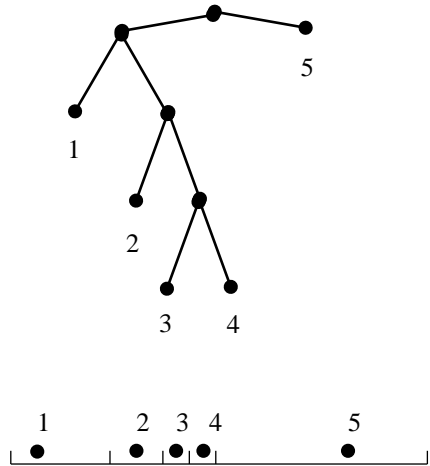}
\caption{ This figure illustrates a tree data structure of
points in 1 dimensions. In this construct, points spatially
close, are also stored nearby. Tree structure is similar
for continuous field, except there is no early termination
of the tree (such as that of number 5 on the figure).}
\label{fig:}
\end{figure}

Finding (joint)
averages like $\avg{\delta_1\ldots \delta_N}$ of a random field $\delta$,
where the $N$-points are taken at a particular configuration,
can be done through summation of $N$-tuples. For the proper
averages, the same algorithm is used for the data, and a random set
describing the geometry of the
survey. For the Monte Carlo estimators an ensemble
of mixed $N$-tuples need to be counted,
but again the same algorithm applies without further complexity.
The algorithm for $N=2$ is described next; higher
orders are exactly analogous, although more tedious to describe.
For details see \cite{MooreEtal2001}
 
The spatial configuration of the two-point function
is characterized by the distance $r$
between the two points: this should lie within a bin,
i.e. $b_1 < r \leq b_2$. An algorithm capable of
calculating the sum $\sum_{r_{12} \leq b} \delta_1 \delta_2$ for any
$b$ yields the answer for a proper bin  $b_1 < r \leq b_2$
as a difference.

\begin{figure} [htb]
\centering
\includegraphics[width=7cm]{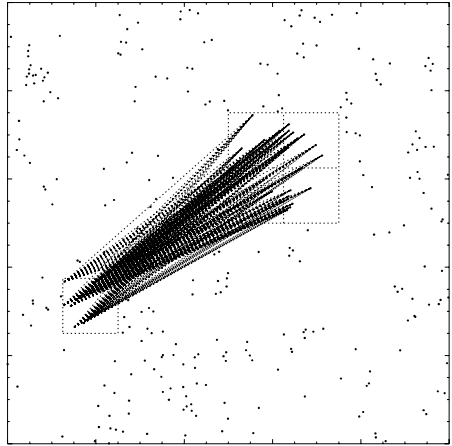}
\caption{ A particular state of the algorithm is
represented. The two squares illustrate two nodes of the
tree, which contain the points inside. The number of points is
cashed, i.e. available at this level of the tree. The pairs of
distances between points are marked with lines.
If case i) or ii) is true (see text),
none of these distances have to be checked,
which in turn furnishes the speed-up of the algorithm.
The smaller squares within the large square represent
the case iii) when the tree has to be split; then
the query is run again recursively.
}
\label{fig:xialg}
\end{figure}

The data points or the pixels of the continuous field can be
arranged in a tree data structure. At every level we store
the number of elements (or values) below,
i.e. sufficient cashed statistics. The sum needed
then can be calculated by a recursive double search on the
tree starting from the root. If the two nodes are $n_1$ and $n_2$
with values $\delta_1$ and $\delta_2$ let us denote with $r$
any distance between points of $n_1$ and $n_2$ (the lines
between the squares of Figure~\ref{fig:xialg}. If we can prove
that i) for any $r > b$
$\rightarrow$ return, or ii)  for any $r \leq b$
$\rightarrow$ add  $\delta_1\times\delta_2$ to the sum and
return, (we are done), else iii) split the tree and continue recursively.
In worst case this procedure descends to the leaves of the tree, but
typically huge savings are realized when whole chunks of data
can be taken into account at coarser levels of the tree.
 
\subsubsection{Algorithms Based on Fourier Transforms}

\begin{figure} [htb]
\centering
\includegraphics[width=10cm]{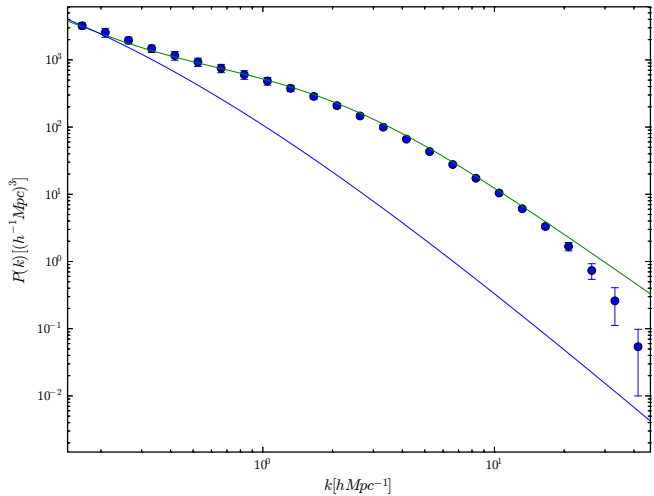}
\caption{ The power spectrum measured in the VLS (Virgo) simulations.
The two-point correlation function have been measured with a
fast hierarchical algorithm, and the edge corrected power spectrum
was obtained from Eq.~\ref{eq:2ptpk}. The errorbars are calculated
from 8 sub-cubes from the simulation, and the two curves show the
linear and non-linear theoretical power spectra \cite{SmithEtal2003}.
The measurement of the correlation function with the Fourier algorithm
is even faster, and produces identical results. 
}
\label{fig:vls2pt}
\end{figure}

As shown next, pair summation can be reformulated to make use of
Fast Fourier Transforms, one of fastest algorithms in existence.
If $P(x) = a_0 + a_1 x + \ldots + a_{n-1}$ is a polynomial,
and $\epsilon = e^{2\pi i/r}$ unit roots, the coefficients
of a discrete Fourier transform of the series $a_i$ can be defined as
\begin{equation}
  \hat a_k = P(\epsilon^k) = a_0+a_1\epsilon^k+\ldots+a_{n-1}\epsilon^{(n-1)k}.
\end{equation}
Direct calculation confirms that $\sum a_k b_{k+\Delta}$ can
be calculated by Fourier transforming the series $a_i$ and $b_i$,
multiplying the resulting $\hat a_k \hat b_k^*$, and finally inverse
Fourier transforming back. This simple observation is the discrete
analog of the Wiener-Khinchin theorem. To obtain correlation
one has to work through subtleties involved with
with the periodic boundary conditions, and multidimensionality.
The final result is a probably fastest algorithm
for calculating correlation functions.
 
The algorithm in broad terms consists of i) placing the points in a
sufficiently fine grid, storing the value $N_{\bf k}$, the number
of objects at (vector) grid point ${\bf k}$, (this step is omitted
if the density field is given, such as CMB maps),
ii) calculating fluctuations of the field by
$\delta = (N-\avg{N})/\avg{N}$, iii) discrete Fourier transform
with a fast FFT engine, iv) multiplying the coefficients v)
Fourier transform back. The same procedure is followed for
the mask with zero padding large enough to avoid aliasing effects.
The resulting inhomogeneous correlation
function is a generalization of the one obtained in the previous
subsections; the usual (homogeneous) correlation function can be obtained
by summing over spherical shells. Edge effect corrected power
spectrum is obtained with yet another Fourier transform. Measurement
of the correlation function on a $768^3$ grid takes about 15 minutes
on one Opteron processor (cf. Figure~\ref{fig:vlsp2t}).

\begin{figure} [htb]
\centering
\includegraphics[width=10cm]{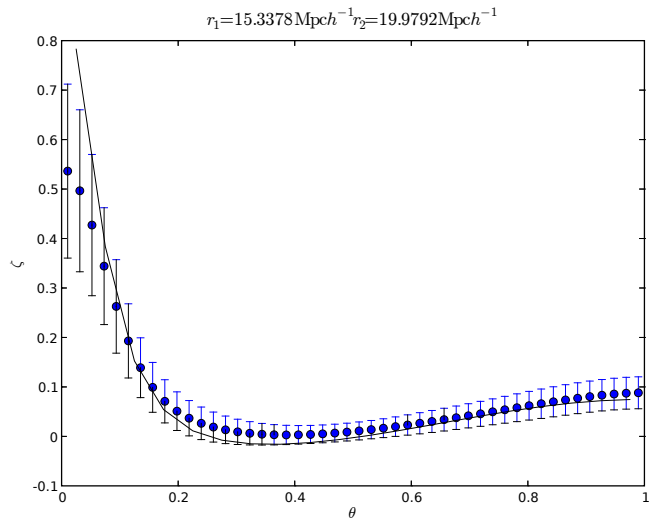}
\caption{The three-point correlation function measured
with a Fourier based algorithm. The data used is the
same as Fig.~\ref{fig:vls2pt}. Theory (solid line) is calculated 
through  Eq.~\ref{eq:b23pt}. Pure leading order perturbation theory
is used, no smoothing effects or non-linearities are taken
into account, which probably account for small differences,
besides cosmic variance. Note that errorbars, estimated
from the same 8 sub-cubes as in the previous figure,  are highly correlated.
}
\label{fig:vls3pt}
\end{figure}

Pure Fourier algorithms are not practical for edge corrected three-point
statistics. However, 
it is possible to combine hierarchical and Fourier methods, to
obtain an algorithm which is faster than either of them.
In Figure~\ref{fig:vls3pt}I show measurements with such an algorithm
in 8 realizations of 260 million particle VLS \cite{JenkinsEtal1998} 
simulations
in a $239.5\mpc$ cube. The measurement using $\simeq 1400$
bins on a $100^3$ 
grid ($\equiv 10^15$ triangles) 
took less than three hours on a 2.4Ghz single CPU. 

%
%



\subsection{Counts in cells algorithms}

I present four different algorithms for counts-in-cells, 
because each of them is optimal for different purposes:
algorithm I. calculates counts in cells with infinite
sampling for small scales, II. is a massively oversampling
algorithm for large scales, III. is a massively oversampling
algorithm for intermediate/small scales, while IV.
is a massively oversampling algorithm for large scales
and arbitrary cell shapes/sizes. II. and IV have
very similar performance, although II. is more
straightforward to generalize for lensing and compensated
filters, while IV. has more flexibility in terms
of cell shape/size. Together I-IV. covers the full
dynamic range of future surveys, with ample overlap
between them for cross check.
 
\subsubsection{CIC I: sweep}
\begin{figure} [htb]
\centering
\includegraphics[width=10cm]{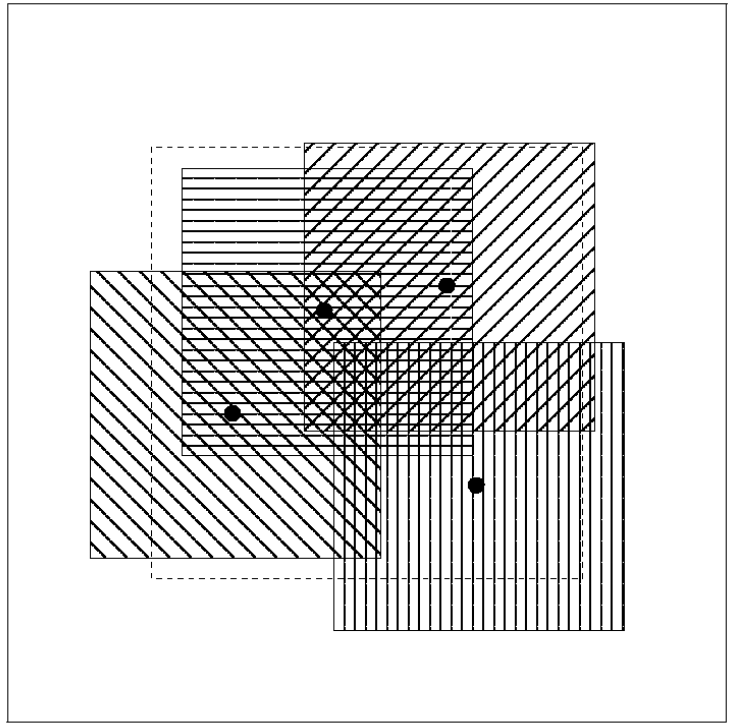}
\caption{Illustrates the geometric calculation of counts in cells.
There are four points within the solid boundary. The centers
of square cells can lie within the dashed boundary. Around
each point a square is drawn to represent the possible centers
of cells which contain that point. The problem of counts
in cells can now be reformulated as calculation of the ratios
of all overlap areas (represented with different shadings
on the figure) within the dashed boundary.
}
\label{fig:sweep}
\end{figure}

This algorithm uses the well known ``sweep'' paradigm from
computational geometry, and it realizes the ideal case of
 $C=\infty$ number of ``random'' cells thrown. Here we
summarize the basic idea in two dimensions only.
 
The geometric interpretation of the probability of finding
$N$ galaxies in a randomly thrown cell is shown on Figure~\ref{fig:sweep}.
There are four points in a rectangular box.sa
Around each object (large dots) a square is drawn, identical to
the sampling cell used for counts in cells.
The possible
centers of random cells all lie within the dashed line, which
follows the boundary of the bounding box.
Since the square around each point corresponds to the possible centers
of (random) cells containing  that same point,
the question can be reformulated in the following way:
let us partition the area of the possible centers of cells
according to the overlap properties of the cells drawn
around the objects. The ratio of the area with $N$ overlaps
to the total area corresponds to $P_N$.
 
Imagine a rigid vertical line moving slowly from the
left of Figure~\ref{fig:sweep}. towards the right;
the boundary can be ignored temporarily.
Before the line touches any of the squares, it sweeps
through an area contributing to $P_0$.
Therefore at the point of first contact all the swept
area contributes to $P_0$ and can be recorded.
After the contact the line is divided
into segments sweeping through areas contributing to $P_0$ and $P_1$
respectively. The boundaries of these segments can be imagined
as two markers on the line, corresponding to the upper and lower
corner the square being touched. As the sweep continues,
the results can be recorded at any contact with the
side of a square during the movement of the line:
the areas swept are assigned according to the markers
on the line to different $P_N$'s.
This is done with a one dimensional sweep on the line
counting the two kinds of
markers. Then the segmentation of the line is updated. Whenever the
line makes contact with the left side of a square, two markers are added,
whenever
it touches the right hand side of a square, the corresponding markers
are dropped.
The boundaries and rectangular masks,
can be trivially taken into account by only starting to record
the result of the sweep when entering the area of possible centers.
Non-rectangular masks can be converted to rectangular by putting
them on a grid.
 
If there are $N$ objects in the plane, the above procedure will
finish after $2 N$ updating.
The algorithm can be trivially generalized for arbitrary rectangles,
any dimensions. For instance in three dimensions the basic
sweep is done with a plane, while
the plane has to be swept by a line  after each contact.
 
From the definition of the algorithm it follows that
the required CPU time scales as $N^D (d/L)^{D(D-1)/2}$ in
$D=2,3$ dimensions, where N is the number of objects, $d/L$ is
the ratio of the scale of the measurement to the characteristic survey length.
While for large scales and large $D$ the algorithm becomes impractical,
it is clear that for small scales it will be quite fast. It
turns out that this is the regime where $C=\infty$ is the most
important \cite{SzapudiColombi1996}.
 
\subsubsection{CIC II: successive convolution}

\begin{figure} [htb]
\centering 
\includegraphics[width=10cm]{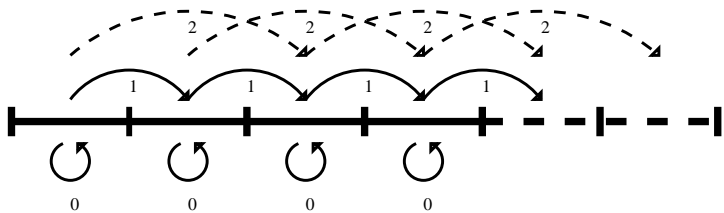}
\caption{ The successive convolution algorithm is illustrated in
one dimension. The solid line represents the grid containing the
data, while the dash line corresponds to auxiliary storage repeating
the first 2 cells. At the 0th level data is placed in the grid,
and counts in cells are calculated. At levels 1 and 2 the solid
and dashed arrows respectively represent summations described in
the text.}
\label{fig:succ_conv}
\end{figure}

This algorithm is essentially a Fourier (or
renormalization) style convolution. It will be
explained in one dimension for simplicity, generalization is obvious.
                                                                                
The computations are performed on the largest possible grid
with $N$ segments
which can be fit into the memory of the computer: this determines
the smallest possible scale $L/N$, where $L$ is the box size,
and $N$ is the base sampling.
A hierarchy of scales are used, with the scale at a given level being
twice the scale at one level lower.
The results one step lower in the hierarchy are used to keep
the number of sampling cells constant even at the largest scales.
Counts in cells can be straightforwardly calculated on the resolution
scale of the grid, i.e. the smallest scale considered. For the calculation
at twice the previous scale the sum of two cells
are always stored in one of the cells, for instance in the one with smaller
index.  Because of the
periodic boundary conditions, auxiliary storage is required to
calculate the sum of the values in the rightmost cell (if the
summations was done left to right), as its right
neighbor is the leftmost cell which was overwritten in the first step.
After these preparatory steps
counts in cells can again be calculated from the $N$ numbers
representing partially overlapping cells. For the next level, twice the
previous scale, one needs the sum of four original resolution cells:
a calculation simply done by summing every other cell of the previous
results into one cell.
At this level, two auxiliary storage spaces are needed because of the
periodicity. In general, at each level in the hierarchy
two cells of the previous results are summed as
a preparatory step, and counts in cells are calculated simply
by computing the frequency distribution of the $N$ sums stored in
the main grid. Auxiliary storage is needed for those rightmost cells,
which have the periodic neighbors on the left end.
 
In D dimensions $2^D$ cells are summed in the preparatory step,
and the auxiliary storage space enlarges the original hypercube.
The scaling of this algorithm is $C \log C \simeq G \log G$
with a preparation step linear in $N$.
 
\subsubsection{CIC III:  tree}
 
This alternative technique for small scales
uses a tree data-structure, similar to the algorithm
defined for the $N$-point correlation functions; it
is explained in three dimensions for convenience.
 
The tree data structure can be thought of as an efficient representation of a
sparse array, since at small scales most of the cells are empty
in a grid spanning the data.
The tree is built up recursively, by always dividing
the particles into two groups based on which half of the volume
they belong to. The same function is called on both halves
with the corresponding particles until
there is no particle in the volume, or the scale becomes smaller
than a predetermined value. At each level the scale and the number
of particles are known, and when an empty volume is reached, all
contained volumes are also empty. These two observations are enough
to insert the book-keeping needed to calculate counts in cells
at all scales while the tree is built. The number of sampling
cells at each level are $2^l$, where $l$ is the level; the original
box is  represented by $l = 0$. Towards smaller scales the
number of cells increases. When $N^3 = 2^l$, where $N$ is the
size of the largest grid of the previous algorithm, the two
techniques should (and do) give the exact same answers. At larger
scales the previous algorithm is superior, since $N > 2^l$,
while this algorithm becomes useful at smaller scales. Just as above,
this procedure can be further improved by shifting the particles
slightly before calculating the tree. However, since this hierarchy
of grids has different numbers of cells, random shifts are more
advantageous. Shifting by a fraction of the smallest scale would not
exhaust the possibilities for any larger scale, while shifting by
a fraction of the largest grid might not shift the underlying grids
at all.
With the introduction of random shifts (oversampling grids),
the dynamic range of algorithms II and III will develop a substantial
overlap, which will be useful for testing. This algorithm also
scales as $N \log N$ (with preset depth limiting).
 
\subsubsection{CIC IV:  cumulative grid }

\begin{figure} [htb]
\centering
\includegraphics[width=7cm]{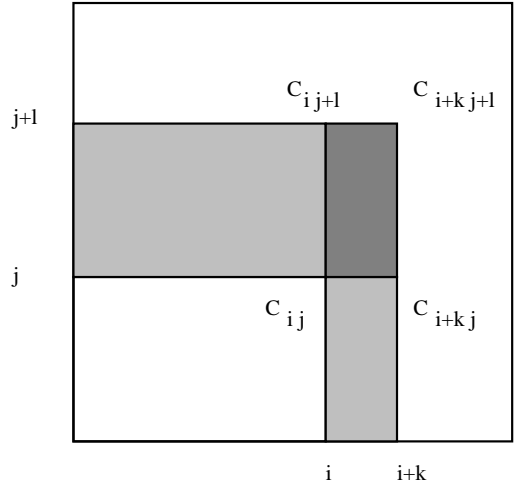}
\caption{ In the cumulative grid algorithm, each grid point is
replaced with a sum of all elements corresponding to rectangles
on the figure, with (0,0) as the lower left, and the grid point
as the upper right coordinate. The dark shaded area is the
required sum, which is calculated as
$c_{ij}+ c_{i+k_1j+k_2}-c_{i+k_1j}-c_{ij+k_2}$ (see text).
}
\label{fig:alg_iv}
\end{figure}

Algorithm II. produces counts in cells results on scales $2^k l$,
where $l$ is the scale associated with the base grid. For
calculating counts-in-cells distribution for arbitrary
scales $[k_1\times l, k_2\times l ]$, the following construction
will be explained in two dimensions; generalization is
obvious.
 
Let us denote the value of the field $n_{ij}$ at the grid-point
$(i,j)$. Let us define another grid, with values
$c_{ij} = \sum_{p \leq i, q \leq j} n_{pq}$. Then the
number of elements in a cell described by $(i,j), (i+k_1,j+k_2)$
can be calculated simply from $c_{ij}+ c_{i+k_1j+k_2}-c_{i+k_1j}-c_{ij+k_2}$
(a well known trick in computational geometry).
The preprocessing is proportional to $N$, and counts in cells
for arbitrary rectangle can be calculated is linear with $C \simeq G$.

\section{Perturbation Theory}

On large scales, where the fluctuations are reasonably small,
clustering of cosmic structures can be understood 
in terms of Eulerian weakly non-linear perturbation theory (PT).
An excellent exposition is found \citep{BernardeauEtal2002} with
substantial reference list. My goal next is to give an
extremely cursory, recipe level introduction to the subject.

PT predicts the behavior of any statistics from Gaussian initial conditions,
assuming an expansion of the density field into first, second, etc.
order, $\delta = \delta^{(1)}+\delta^{(2)}+\ldots$. This assumption,
when substituted into Euler's equations for the cosmological
dark matter ``fluid'', yields a set of consistent equations
for the different orders. The resulting perturbative expansion is
most conveniently expressed in Fourier space with kernels $F_N$ as,
\begin{equation}
   \delta^{(N)}(k) = \int d^3q_1\ldots d^3q_N F_N(q_1,...,q_N) 
\delta_D(q_1+\ldots+q_N-k) \delta^{(1)}(q_1)\ldots \delta^{(1)}(q_N)
\end{equation} 
and an analogous equation for the velocity potential. Euler's equations
lead to a simple, albeit ``combinatorially exploding'' recursion
for the kernels \cite{GoroffEtal1986}. 
The most important kernel is the first non-trivial $F_2$
\begin{equation}
F_2({\bf q_1},{\bf q_2}) = 
1+\mu+(\frac{q_1}{q_2} +\frac{q_2}{q_1})\cos(\theta)+
(1-\mu)\cos(\theta)^2,
\end{equation} 
where $\theta$ is the angle between the two vectors (in this equation
explicitly denoted by bold face), and $\mu = \frac{3}{7}\Omega^{-1/140}$
\cite{KamionkowskiBuchalter1999}.

The leading order calculation of e.g., any third order statistic can
be obtained simply by $\avg{\delta^3}_c = 
\avg{(\delta^{(1)}+\delta^{(2)}+\ldots)^3}_c = 
3 \avg{(\delta^{(1)})^2\delta^{(2)}} + \rm{higher\, orders}$. For Gaussian
initial conditions, this leads to formula in terms of $P(k)$, the
initial (or linear) power spectrum, and the $F_2$ kernel.

When the above expansion is executed carefully PT predicts
\cite{Fry1984a,GoroffEtal1986,BouchetEtal1995,HivonEtal1995}
that the bispectrum in the weakly non-linear regime is
\begin{equation}
  \left\lbrace\left(\frac{4}{3}+\frac{2}{3}\mu\right)P_0(x)+
  \left(\frac{k_1}{k_2}+\frac{k_2}{k_1}\right)P_1(x)+
  \frac{2}{3}\left(1-\mu\right)P_2(x)\right\rbrace P(k_1)P(k_2)+\rm{perm.}
\end{equation}
where $x$ is the cosine of the angle between the two wave vectors, $P_l$
are Legendre polynomials, and $P(k)$ is the linear power spectrum. 
We have written the $F_2$ kernel
in terms of Legendre polynomials, to make it explicitly
clear that the first permutation depends only on terms up to quadrupole.

Together with the formulae in section~\ref{sec:sym} this fact
can be used for the prediction of the first permutaion
of three-point correlation
function in the weakly nonlinear regime:
\begin{eqnarray}
   \xi^3_0(r_1,r_2) =&  \int\frac{k_1^2}{2 \pi^2}dk_1 
    \frac{k_2^2}{2 \pi^2} dk_2 \left(\frac{4}{3}+\frac{2}{3}\mu\right)) 
     P(k_1)P(k_2)j_0(k_1r_1)j_0(k_2r_2) 
    \nonumber\\
   \xi^3_1(r_1,r_2) =& -\int\frac{k_1^2}{2 \pi^2}dk_1 
    \frac{k_2^2}{2 \pi^2} dk_2 \left(\frac{k_1}{k_2}+\frac{k_2}{k_1}\right)
    P(k_1)P(k_2)j_1(k_1r_1)j_1(k_2r_2) 
    \nonumber\\
   \xi^3_2(r_1,r_2) =&  \int\frac{k_1^2}{2 \pi^2}dk_1 
    \frac{k_2^2}{2 \pi^2} dk_2 \frac{2}{3}\left(1-\mu\right)
    P(k_1)P(k_2)j_2(k_1r_1)j_2(k_2r_2)  
   \label{eq:b23pt}
\end{eqnarray}
I emphasize that the above equation corresponds to the first permutation of the
perturbation theory kernel: the other permutations are dealt
with the same way, and the results are added.
All the integrals factor into one dimensional ones, which simplifies
the calculation in practice.
It can be shown with tedious calculations
that the above formulae are equivalent
to \cite{JingBorner1997}. These equations has been used for
the predictions of Figure~\ref{fig:vls3pt}. One has to be somewhat
careful to integrate the Bessel functions with sufficient
accuracy.  

For completeness we present the analogous formulae for the
projected three-point correlation function (see \S 9), 
which is integrated over line of sight coordinates to avoid the effects of
redshift distortions. The perturbation
theory kernel is simply rewritten in Fourier modes \cite{Zheng2004}
\begin{equation}
  \left\lbrace\left(\frac{3}{2}-\frac{1}{2}\mu\right)+
  \left(\frac{k_1}{k_2}+\frac{k_2}{k_1}\right)cos(\theta)+
  \frac{1}{2}\left(1-\mu\right)cos(2\theta)\right\rbrace P(k_1)P(k_2)+
\rm{perm.}
\end{equation}
which yields the equivalent result for the projected three-point
function $\zeta^p$ as
\begin{eqnarray}
   \zeta^p_0(r_1,r_2) =&  \int\frac{k_1}{2 \pi}dk_1 
    \frac{k_2}{2 \pi} dk_2 \left(\frac{3}{2}-\frac{1}{2}\mu\right)) 
     P(k_1)P(k_2)J_0(k_1r_1)J_0(k_2r_2) 
    \nonumber\\
   \zeta^p_1(r_1,r_2) =& -\int\frac{k_1}{2 \pi}dk_1 
    \frac{k_2}{2 \pi} dk_2 \left(\frac{k_1}{k_2}+\frac{k_2}{k_1}\right)
    P(k_1)P(k_2)J_1(k_1r_1)J_1(k_2r_2) 
    \nonumber\\
   \zeta^p_2(r_1,r_2) =&  \int\frac{k_1}{2 \pi}dk_1 
    \frac{k_2}{2 \pi} dk_2 \frac{1}{2}\left(1-\mu\right)
    P(k_1)P(k_2)J_2(k_1r_1)J_2(k_2r_2) 
   \label{eq:b23pt2d}
\end{eqnarray}
Here both $r_i$ and $k_i$ are two-dimensional vectors perpendicular
to the line of sight, and the other two permutations have to
be added up as previously.

\section{Bias}

Most observervations record the distribution of ``light'' at some wavelength
(except some still singular cases, such as neutrino astronomy, 
or air shower detectors), 
might have spatial distribution different from that of the
underlying dark matter or ``mass''. This  difference
means that our estimators might give us a ``biased'' view
of the statistics of the underlying dark matter. This bias
is caused by the complex interplay of non-linear gravitational
dynamics with dissipative physics. While there have already been
advances in ab initio modeling of this complicated 
process, the most fruitful approach is still phenomenological.

Since galaxy formation is a stochastic process, we can imagine
that there is no one-to-one correspondence between dark matter
and galaxy density, the former only determines the probability
of forming a galaxy (together with its environment and merger
history). This general case is called {\em stochastic } biasing
\cite{DekelLahav1999,ScherrerWeinberg1998},
while, if there is a functional dependence between the two fields,
it is {\em deterministic}. In mock catalogs created using 
semi-analytic galaxy formation models e.g., \cite{BensonEtal2000},
most of the stochasticity of the bias is simply due to shot noise
\cite{SzapudiPan2004}.

If there is a functional relationship between the dark matter
and galaxy density fields, the simplest possible relation is
{\em linear}, the more general ones are {\em non-linear}. 

Linear bias is described by the equation of $\delta_g = b\delta_{DM}$,
where $b$ is a constant. Note that this is an inconsistent equation
for $b > 1$  yielding the minimum value of the galaxy density
field $-b < -1$, obviously non-sense. Yet, owing to its simplicity,
this is the single most used bias model. 

A non-linear generalization contains an arbitrary function
$\delta_g = f(\delta_{DM})$. This function can either
be assumed to be a general function with a few parameters
(e.g. exponential), or can be expanded. The two most
used expansions are Taylor expansion, or expansion into
Hermite-polynomials (and thereby expanding around a Gaussian).

In the weakly non-linear regime, the coefficients of the
Taylor expansion are the non-linear bias coefficients;
\begin{equation}
   \delta_g = f(\delta_{DM}) = \sum_N \frac{b_N}{N!} \delta_{DM}^N.
\end{equation}
This equation can be used perturbatively to calculate the moments
of the biased (galaxy) density field in terms of the moments
of the underlying field, and bias coefficients 
\cite{Szapudi1990,FryGaztanaga1993,Matsubara1995}.
In typical applications, the functional relationship is set up between
smoothed fields: it is important to note that smoothing and
biasing are two operations which do not commute. Therefore
the bias coefficients have a somewhat obscure physical meaning,
as they depend not only on the physics of galaxy formation,
but on the nature of smoothing itself. Note also, that the
zeroth coefficient $b_0$ is set to ensure $\avg{\delta_g} = 0$,
i.e. it is non-zero.

Symmetries can be used used to construct estimators which constrain the bias
in the weakly non-linear regime.
To second order, the biased reduced bispectrum transforms as $q = Q/b+b_2/b^2$
\cite{Fry1994}, where the lower case denotes the galaxy (measured),
and the upper case the dark matter (theory) values.
It is clear that $b_2$ can only effect the monopole term.
Thus a simple estimator for the bias can be constructed as 
\cite{Szapudi2004a}
\begin{eqnarray}
   b =& \frac{Q_1}{q_1}=\frac{Q_2}{ q_2}\cr
   b_2 =& q_0b^2-Q_0b.
\end{eqnarray}
According to the equations, the  quadrupole to dipole ratio
does not depend on the bias, thus it serves as a novel, useful
test of the underlying assumptions:
a quasi-local perturbative, deterministic bias model and perturbation theory.
Figure 4. shows the dipole to quadrupole ratio for BBKS,
and EH power spectra, respectively.
The range of $k$'s to be used for bias extraction can be determined from
contrasting the measurements  with these predictions. Note that
scales where baryon oscillations are prominent are barely accessible
with present redshift surveys. On smaller scales non-linear evolution
is likely to modify these prediction based purely on leading order
perturbation theory (\cite{MeiksinEtal1999}.
\begin{figure}[htb] 
\centerline{
\includegraphics[width=6cm]{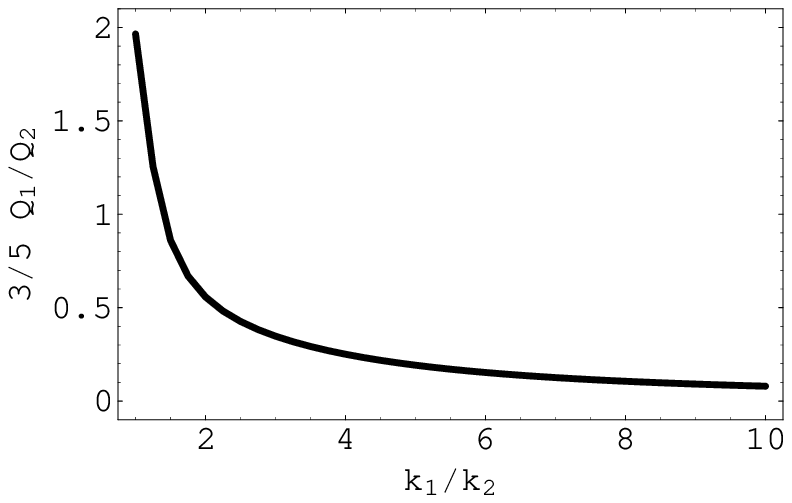}
\includegraphics[width=6cm]{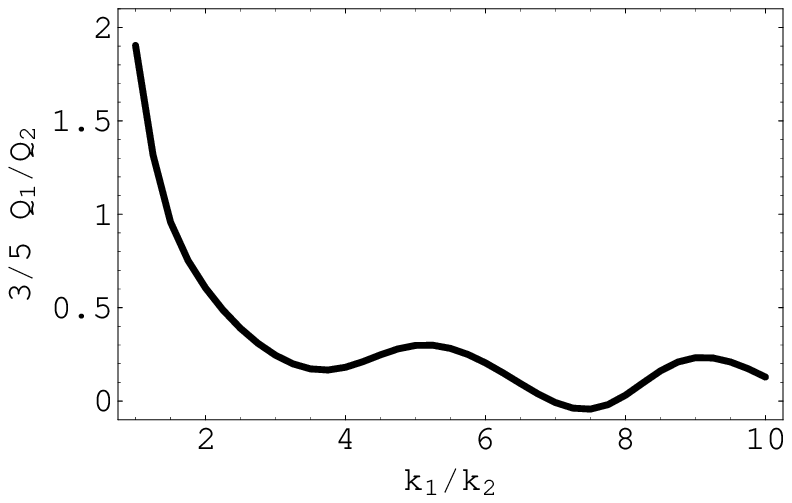}}
\caption{Left: the dipole to quadrupole  ratio $3/5 Q_1/Q_2$
is plotted for a BBKS theory (thick solid line) for $k=0.01$.
The right panel shows the same using the EH fit for comparison,
featuring baryonic oscillations.}
\end{figure}

The above simple theory illustrates that three-point statistics
can constrain the first two bias parameters in the weakly non-linear
regime. Similarly four point statistics constrain the first three
parameters, etc.

An alternative to the above simple theory, and possibly more
useful on smaller scales, is the halo model. The physical picture
is simple \cite{SoneiraPeebles1977}: we try
to model the distribution by some entities, ``halos'', which
have a relatively simple distribution (from perturbation theory).
Large scale correlations thus follow from the ``halo-halo'' type 
terms, while all the small scale statistics follows from the halo
profile (``1-halo term''), and the distribution of halo masses. In these
notes we cannot give justice to halo models, the interested
reader is referred to \cite{CooraySheth2002}, and references
therein. Halo models have
considerable success in describing the two-point statistics
of the dark matter distributions, and they provide a reasonable
approximation (currently at the 20\% level) to the three-point
statistics. Galaxy bias is then described by the halo occupation
probabilities. 

For halo modell predictions, higher order Bessel integrals
are needed \cite{FosalbaEtal2005},  
but the principle is the exactly the same as
Eq.~\ref{eq:b23pt} (see also \cite{TakadaJain2003} for a different
approach). While halo models are physically motivated, 
the best fit parameters differ for two and three-point
statistics  thus the physical
meaning of the parameters is still somewhat nebulous.

\section{Redshift Distortions}

Redshift distortions, together with biasing,
represent the most uncertain aspect
of the phenomenology of higher order statistics. 
Here I only offer a superficial overview of this 
complicated topic; for a comprehensive review
of redshift distortions on two-point statistics,
see \cite{Hamilton1998}

The physical idea is fairly simple: in redshift surveys,
the radial distance to an object is deduced from its
redshift using the Hubble relation in our expanding
universe. Therefore, the ``peculiar velocity'', the
velocity of an object with respect the the average Hubble
flow, will add to the measured distance. Such a space,
where position of an object is given in
a spherical coordinate system with two angles, and
a radial distance which contains (a random) velocity
is called {\em redshift space}, as opposed to real
space (the one we would really want to measure).

The spatial distribution of objects will no longer
be translation invariant, although it will still have
rotational symmetry around the observer. The deviation
from translation invariance is of the observed distribution
is called ``redshift distortion''. Qualitatively, it is
mainly due to two distinct effects: on small scales,
velocity dispersion of virialized clumps will look
elongated along the line of sight: popularly called
the ``finger of god'' effect. On larger scales, infall
velocities cause distortions perpendicularly to the
line of sight, the Kaiser-effect. 

The breaking of translational symmetry means that 
the redshift-space quantities depend on larger
number of parameters, e.g. the two- and three-point
correlation functions will depend on 3 and 6
parameters respectively (2, and 5 in the distant
observer approximation).

Redshift distortions can be taken into account in perturbation
theory. In the distant observer approximation, when all
lines of sights can be take to be parallel, the
effect has been calculated for
the three-point function by \cite{ScoccimarroEtal1999a}.
The general case needs a complex multipolar expansions;
the results are fairly tedious and will be presented
elsewhere.

Besides predicting the redshift space quantities using
theory or simulation, on can sweep redshift distortions
under the rug by  introducing the
``projected'' correlation functions. These assume
a distant observer approximation, where one can consider
the $N$-vectors to pointing from the observer to $N$
vertices of an $N$-point configuration parallel. Then,
one can introduce $\pi-\sigma$ coordinates \cite{Peebles1980},
with $\pi_i$ vectors parallel to the line of sight, while
the $\sigma_i$ vectors perpendicular to it. 

The important point about this parametrization is that, approximately,
only $\pi$ coordinates are affected by redshift distortions. A
convenient, redshift distortion free quantity is obtained by
integrating over the redshift direction, i.e. $\pi$ coordinates.
The resulting object, the ``projected $N$-point correlation
function'', integrated over $N-1$ $\pi$ coordinates is free 
of redshift distortions. Although it is similar to angular
(projected) correlation function, the units of the $\sigma$
coordinates are still $\mpc$, and, if properly done, its
signal to noise is higher. 

Fourier space analogs of the  above idea use $k_\bot$
and $k_\|$ for perpendicular and parallel $k$ vectors.
E.g., the redshift space bispectrum is parametrized by the five parameters
$B(k_{1,\bot},k_{2,\bot},k_{3,\bot},k_{1,\|},k_{2,\|})$,
with $\bot$ denoting transverse, and $\|$ parallel quantities
with respect to the line of sight in the distant observer
approximation. Interestingly, the real space bispectrum can be estimated
from taking $k_{1,\|}\simeq k_{2,\|}\simeq 0$ \cite{Szapudi2004a}.

\section{Example: Conditional Cumulants}

In my lectures, I presented an detailed example
using a new set of statistics, conditional cumulants, 
closely following \cite{PanSzapudi2004}.
It illustrates most of the generic 
features of the theory, baring its strengths and weaknesses.

Conditional cumulants represent
an interesting and sensible compromise between $N$-point correlation
functions and cumulants measured from moments of counts in cells.
As we will see next, they can be understood as 
degenerate $N$-point
correlation functions, or integrated monopole moments of the bispectrum,
and they are closely related to neighbor counts.
They share accurate edge corrected estimators
with $N$-point correlation functions,
yet, they are as straightforward to measure and interpret
as counts in cells. 

\subsection{Basics}

The general conditional cumulants, Eq.~\ref{eq:cc}, have been defined 
as the joint connected 
moment of one unsmoothed and $N-1$ smoothed density fluctuation
fields.  In the most general case, each top hat window
might have a different radius.
Further simplification arises if all the top hats are the same, i.e.
we define 
$U_N(r)$ with $r_1=\ldots =r_{N-1}=r$ as the degenerate conditional cumulant
c.f. \citep{BonomettoEtal1995}. $U_N$ subtly differs from the 
usual cumulant of smoothed field $\overline{\xi}_N$ 
by one less integral over the window function.

The second order, $U_2$, is 
equivalent to the (confusingly named) $J_3$ integral
e.g., \citep{Peebles1980},
\begin{equation}
U_2(r)=\frac{3}{r^3}J_3(r)= \frac{1}{(2\pi)^3}\int P(k) w(kr) 4 \pi k^2 dk\ , 
\end{equation}
where $w(kr)=3(\sin kr-kr \cos kr )/(kr)^3$ is the Fourier transform
of $W_r(s)$, and $P(k)$ is the power spectrum. 

For higher orders, we can construct
reduced conditional cumulants as
\begin{equation}
R_N(r)=\frac{U_N(r)}{U_2^{N-1}(r)}\ .
\end{equation}

Both $U_N$ and $R_N$ have deep connection with moments
of neighbor counts e.g., \citep{Peebles1980} as we show next. 
Let us define the partition function
$Z[J] = \avg{\exp({\int i J \rho})}$ c.f., \citep{SzapudiSzalay1993},
where $\rho$ is the smooth density field.
Then we can use the special source function 
$i J(x) = W(x) s+ \delta_D(x) t$ to obtain the
generating function $G(s,t)$. This is related to the
generating function of neighbor counts factorial moments as
$G(s) = \partial_t G(s,t) |_{t = 0}$. Explicitly,
\begin{equation}
  G(s) =   \sum_{M \ge 0} \frac{(s n v)^M}{M!} U_{M+1}\,
        \exp{\sum_{N \ge 1} \frac{(s n v)^N}{N!} \overline{\xi}_N}\ ,
\end{equation}
where $n v = \bar N$ is the average count of galaxies,
and $\overline{\xi}=U_1 =1$ by definition. This generating
function can be used to obtain $U_N$'s and/or $R_N$'s 
from neighbor count
factorial moments analogously as the generating functions
in \cite{SzapudiSzalay1993a} for obtaining $S_N$'s from
factorial moments of counts in cells. For completeness,
the generating function for neighbor counts' distribution
is obtained by substituting $s \rightarrow s -1$, while
the ordinary moment generating function by $s \rightarrow e^s-1$.
Expanding $G(e^s-1)$ recovers the formulae in 
\cite{Peebles1980}, \S 36. The above generating
function allows the extraction of $U_N$ from neighbor count
statistics for high $N$. The situation is analogous to
the CIC theory presented in \cite{SzapudiSzalay1993a},
and it is fully applicable to neighbor counts statistics
with minor and trivial modifications.

So far our discussion has been entirely general;
in what follows we will focus on $N=3$, i.e. the first non-trivial
conditional cumulant $U_3$.
$U_3(r_1, r_2)$ is directly related to bispectrum by
\begin{eqnarray}
U_3(r_1, r_2)& =\frac{1}{(2\pi)^6}\int B({\bf k}_1, {\bf k}_2, {\bf k}_3) 
                \delta_D({\bf k}_1+{\bf k}_2+{\bf k}_3) \nonumber \\
             & w(k_1r_1) w(k_2 r_2)d^3 k_1 d^3 k_2 d^3k_3\ ,
\end{eqnarray}
where $\delta_D$ is the Dirac delta function.
To further elucidate this formula, we use the multipole
expansion of bispectrum and 
three point correlation function  of Eqs.~\ref{eq:3ptbl3d} and
\ref{eq:transform3D} to find
\begin{eqnarray}
U_3(r_1, r_2)=&\frac{4 \pi}{V_1 V_2}\int_0^{r_1}\int_0^{r_2} 
\zeta_0(r_1, r_2) r_1^2 r_2^2 dr_1 dr_2\\
=&
\frac{4\pi}{(2\pi)^6}\int dk_1 dk_2 \frac{3k_1}{r_1}
\frac{3k_2}{r_2} j_1(k_1 r_1) j_1(k_2 r_2) B_0(k_1, k_2)\ ,
\end{eqnarray}
in which $j_1$ is the first order spherical Bessel function. We see
the $U_3$ depends only on the monopole moment of the bispectrum/three-point
correlation function. This property significantly simplifies the
transformation of this statistic under redshift distortions.

\subsection{$U_3$ in the Weakly Non-linear Perturbation Regime}

Using the general machinery of perturbation theory, 
one can predict the conditional cumulants to leading order
as matter of simple calculation. For third order, the
results, as usual, will depend on the $F_2$ kernel, and the
linear power spectrum (or correlation function). We leave
as an exercise for the reader to show that
\begin{eqnarray}
R_3 (r_1,r_2)\equiv &\frac{U_3(r_1, r_2)}{U_2(r_1) U_2(r_2)}=
\frac{34}{21} \left[ 1+
\frac{\overline{\xi}(r_1, r_2)}{U_2(r_1)} +\frac{\overline{\xi}(r_1, r_2)}{U_2(r_2)} \right] \nonumber \\
 + \frac{1}{3}&\frac{\overline{\xi}(r_1,r_2)}{U_2(r_1)} \left[ \frac{d \ln U_2(r_2)}{d\ln r_2} +
\frac{\partial \ln \overline{\xi}(r_1,r_2)}{\partial \ln r_2} \right] \\
 + \frac{1}{3}&\frac{\overline{\xi}(r_1,r_2)}{U_2(r_2)} \left[ \frac{d \ln U_2(r_1)}{d\ln r_1} +
\frac{\partial \ln \overline{\xi}(r_1,r_2)}{\partial \ln r_1} \right] 
\nonumber \ ,
\end{eqnarray}
in which $\overline{\xi}(r_1, r_2)=\frac{1}{2\pi^2}\int k^2 P(k) W(kr_1) W(kr_2)dk$.
The special case when $r_1=r_2=r$ reads
\begin{equation}
R_3=\frac{34}{21}\left[ 1+ 2\frac{\sigma^2}{U_2}\right] +\frac{1}{3}\
\frac{\sigma^2}{U_2}\left[ 2 \frac{d\ln U_2}{d \ln r}+ \frac{d \ln \sigma^2}{d \ln r}
\right]\ ,
\label{eq:r3}
\end{equation}
where $\sigma^2=\frac{1}{2\pi^2}\int k^2P(k)W^2(kr)dk$.
Note the similarity of $R_3$ with the
skewness, which is calculated in weakly non-linear
perturbation theory as  $S_3 = 34/7- d \ln \sigma^2/d \ln r$
\citep{JuszkiewiczEtal1993,Bernardeau1994c}.

\subsection{Measurements of Conditional Cumulants in Simulations}
%
%
%
%

%
%
\begin{figure}
\centerline{
\includegraphics[width=6.5cm]{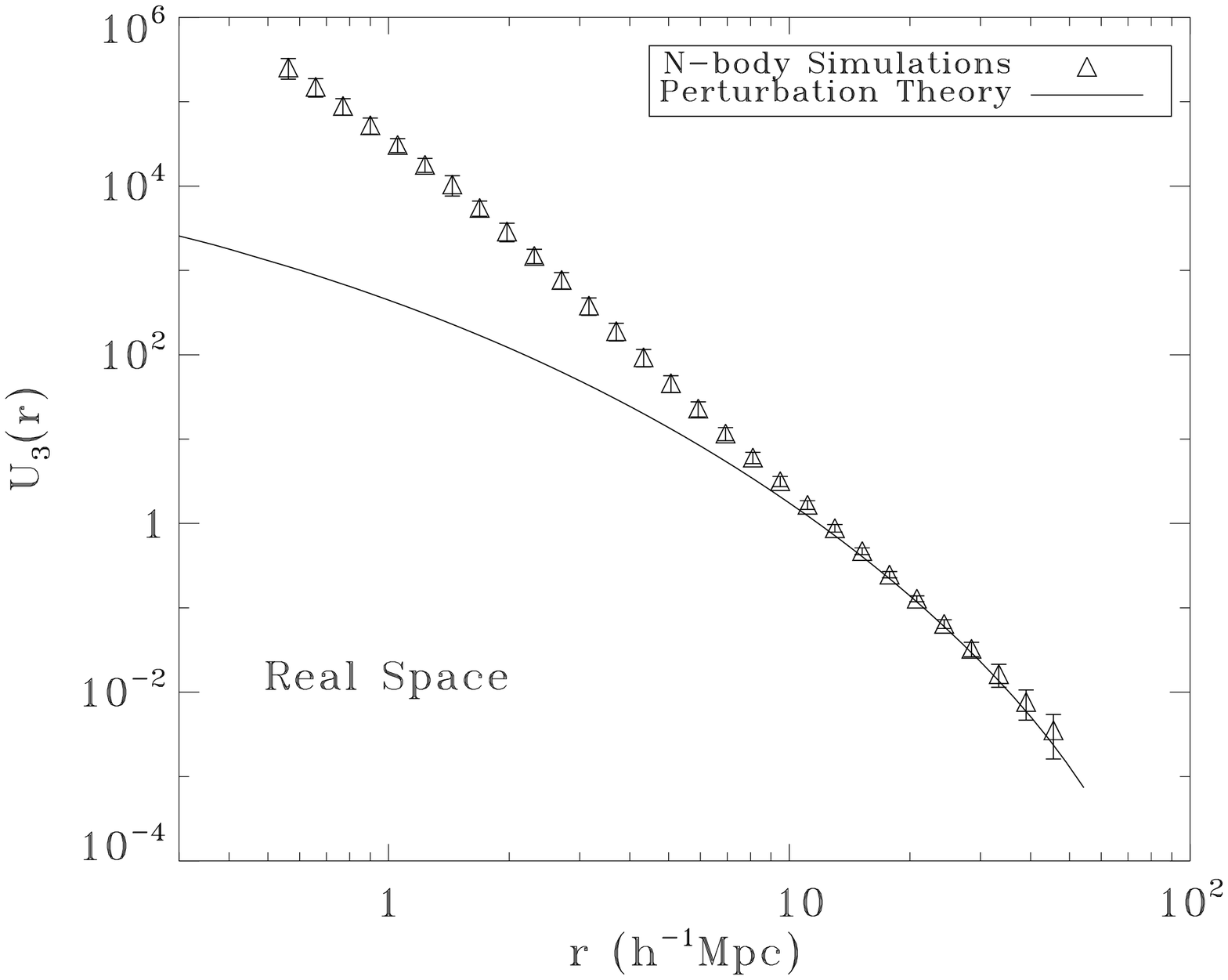}
\includegraphics[width=6.5cm]{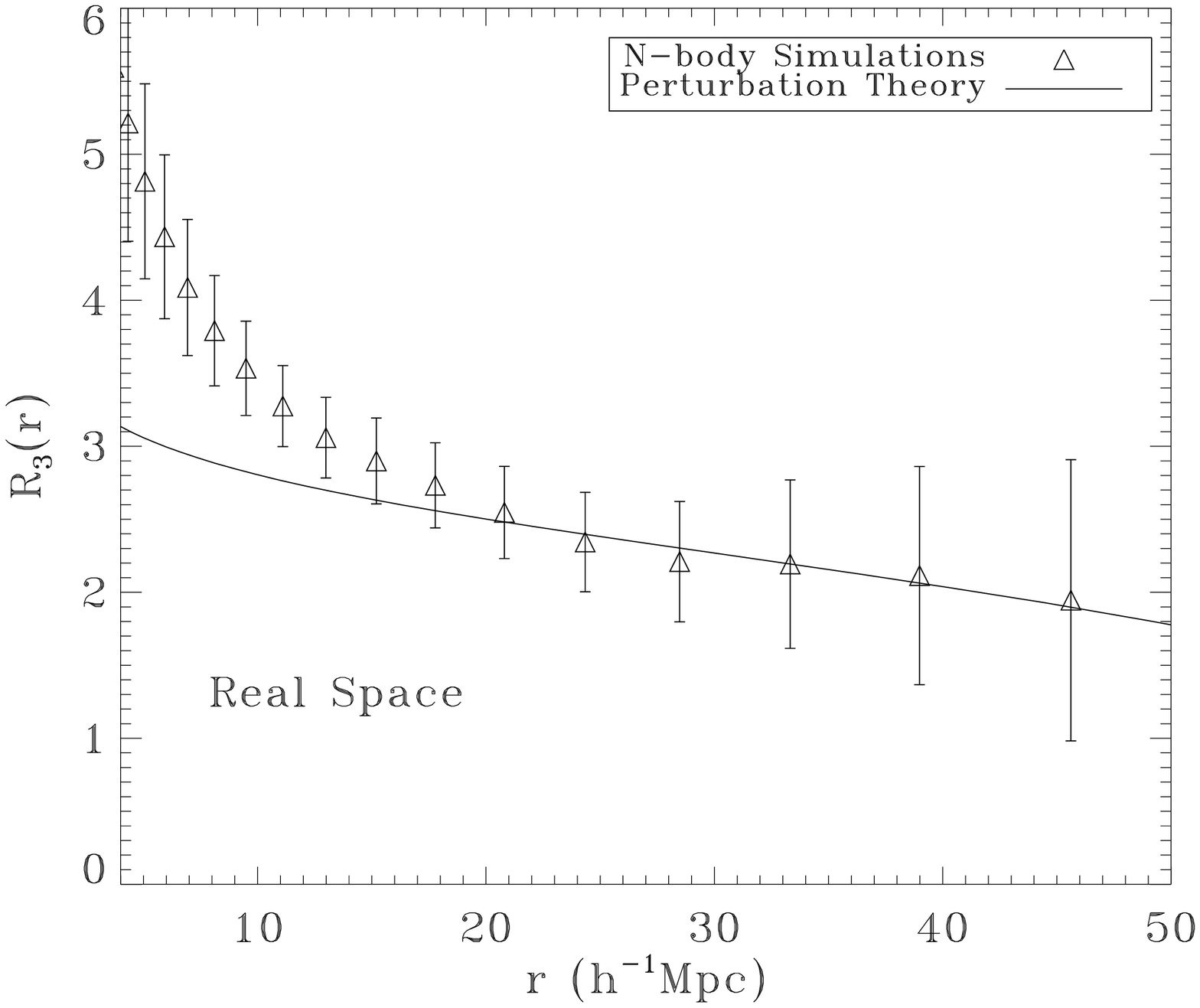}}
\caption{ Predictions of weakly non-linear perturbation
theory for third order conditional cumulant
$U_3$ (left), and the reduced statistics $R_3$ (right) are 
in real space (solid line) are
compared with measurements in N-body simulations
(triangles with errorbars) with $\Lambda$CDM cosmology.
The agreement is remarkable above $\gtrsim 20\mpc$}
\end{figure}
$U_n(r)$ can be measured similarly to 
$N$-point correlation functions. For instance $U_2$ can be thought
of as a two point correlation function
in a bin $[r_{lo}, r_{hi}] \equiv [0,r]$. Taking
the lower limit to be a small number instead of 0, one can
avoid discreteness effects due to
self counting (this is equivalent to using factorial moments
when neighbor counts are calculated directly). 
Given a set of data and random points,
the class of estimators of \ref{eq:ssnpt} provides an edge (and
incompleteness) corrected technique to measure conditional
cumulants. Existing
$N$-point correlation function codes can be used for the estimation;
for higher then third order, one also has to take connected moments
in the usual way.

While the above suggests a scaling similar to $N$-point correlation
functions, the relation to neighbor count factorial
moments outlined in the previous section 
can be used to realize the estimator using two-point
correlation function codes. To develop such an estimator, 
neighbor count factorial
moments need to be collected for each possible combinations,
where data and random points play the role of center and neighbor.

Note that the edge correction of
Eq.~(\ref{eq:ssnpt}) is expected to be 
less accurate for conditional cumulants, than for $N$-point
correlation functions, however, the estimator will be
more accurate than CIC estimators. Several alternative
ways for correcting edge effects are known, which
would be directly applicable to conditional cumulants
\citep{Ripley1988, KerscherEtal2000, PanColes2002}. In what
follows, we use  Eq.~(\ref{eq:ssnpt}) for all results presented. 


To test PT of the conditional cumulants, 
we performed measurements in $\Lambda$CDM 
simulations by the Virgo Supercomputing Consortium \cite{JenkinsEtal1998}. 
We  used outputs of the Virgo simulation and the VLS (Very Large Simulation).
Except for box sizes and number of particles, 
these two simulations have identical 
cosmology parameters: $\Omega_m=0.3$, 
$\Omega_v=0.7$, $\Gamma=0.21$, $h=0.7$ and
$\sigma_8=0.9$. In order to estimate measurement errors, 
we divide the VLS simulation into
eight independent subsets each with the same size and geometry 
the original Virgo simulation. In total, we have used
the resulting nine realizations to estimate errors. 
Note that we corrected for cosmic bias by always
taking the average before ratio statistics were formed.

Our measurements of the second and third order conditional cumulants 
are displayed in Figs. 1 and 2, respectively. Results from EPT (Eq.~\ref{eq:r3}
are denoted with solid lines. The measurements in simulations are
in excellent agreement with EPT, especially on
on large scales $\gtrsim 20\mpc$. 

\subsection{Redshift Distortions}

As three-dimensional galaxy catalogs are produced inherently in redshift space,
understanding effects of redshift distortions on
these statistics is crucial before practical applications can follow.
In the distant observer approximation, the formula
by \cite{Kaiser1987,LiljeEfstathiou1989} is expected
to provide  an excellent approximation for $U_2(r)$. 
According to \S 2, we only need to consider
the monopole enhancement 
\begin{equation}
U_2(s)=\left( 1+\frac{2}{3}f+\frac{1}{5}f^2 \right) U_2(r)\ ,
\end{equation}
where $f\approx \Omega_m^{0.6}$. This formula essentially predicts
a uniform shift of the real space results. To test it, we repeated our
measurements in redshift space, and found that the above is indeed
an excellent approximation in redshift space.

Considering the relatively simple, monopole nature
of the statistics, we expect that the overall effect on $U_3$ should
also be a simple shift, similarly to the Lagrangian calculations by
\cite{HivonEtal1995} and the more general Eulerian 
results by \cite{ScoccimarroEtal1999a}. 
Specifically, we propose that ratio of
$R_3$ in redshift space to that in real space can be approximated by
\begin{equation}
\frac{ 5(2520 + 3360 f+1260 f^2+9 f^3-14 f^4)}{98(15+10f+3f^2)^2} \cdot 
\frac{7}{4}\ \ \ .
\label{eq:bm}
\end{equation}
This is motivated by the notion that the shift from redshift distortions
of equilateral triangles should be similar to the corresponding
shift for our monopole statistic. Our simulations results (see Fig.~3)
show that this simple idea is indeed a surprisingly good approximation, 
although the phenomenological theory based on the above formula appears to
have $\simeq 5\%$ bias on scales $\gtrsim 20\mpc$ where
we expect that weakly non-linear perturbation theory is a good
approximation. For practical applications, this bias can be calibrated
by $N$-body, or 2LPT \citep{Scoccimarro2000} simulations.

In addition to the above simple formula, we have calculated the
shift due to redshift distortions 
by angular averaging the bispectrum monopole term in 
\cite{ScoccimarroEtal1999a}.
We have found that the results over-predict redshift distortions,
however, they would agree with simulations at the 1-2\% level
if we halved the terms classified as FOG (finger of god).
At the moment there is no justification for such a fudge
factor, therefore we opt to use the above phenomenology, which
is about 5\% accurate.
While redshift distortions of third order statistics 
are still not fully understood due to the non-perturbative nature of the
redshift space mapping (R. Scoccimarro, private communication),
detailed calculations taking into account velocity dispersion
effects will improve the accuracy of the redshift space theory $U_3$.

\begin{figure}
\centerline{\includegraphics[width=10cm]{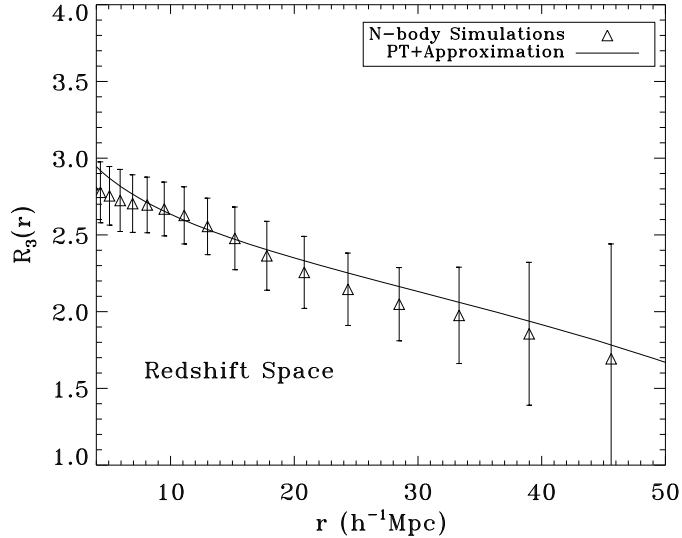}}
\caption{Same as Figure 2, lower panel, but for
$R_3$ in redshift space. The solid line shows
phenomenological model based on Equation~\ref{eq:bm}. The
theory appears to be a reasonable approximation at the 5\% level.}
\end{figure}

For applications to constrain bias, one has to keep in mind that redshift
distortions and non-linear bias do not commute. However, at
the level of the above simple theory, it is clear that one can understand
the important effects at least for the third order statistic.
There are several ways to apply conditional cumulants for bias
determination, either via combination with another statistic
\citep{Szapudi1998a}, or using
the configuration dependence of the more general $R_3(r_1,r_2)$.
One also has to be careful that in practical
applications ratio statistics will
contain cosmic bias \citep{SzapudiEtal1999b}. We propose that joint
estimation with $U_2$ and $U_3$ will be more fruitful, even
if $R_3$ is better for plotting purposes. Details of the techniques
to constrain bias from these statistics, as well we determination
of the bias from wide field redshift surveys is left for future work.

Another way to get around redshift distortions is to
adapt conditional cumulants for projected and angular quantities. 
Such calculations are straightforward, 
and entirely analogous to those performed for $S_3$ in the
past. Further possible generalization of our theory would be to use
halo models \citep{CooraySheth2002} to extend the range
of applicability of the theory well below $20\mpc$. 
These generalizations are left for subsequent research. 

\section{ Acknowledgments}

This work was supported by NASA through grants AISR NAG5-11996,
ATP NAG5-12101, and by NSF through grants AST02-06243 and
ITR 1120201-128440, as well by Valencia summer school,
one of the best organized and most enjoyable ever.
The author would like to thank Stephane Colombi,
Pablo Fosalba, Jun Pan, Simon Prunet, Alex Szalay for many discussions
and contributions.
The simulations used have been carried out by the Virgo
Supercomputing Consortium using computers based at Computing Centre of
the Max-Planck Society in Garching and at the Edinburgh Parallel
Computing Centre\footnote{The data are publicly available at
{http://www.mpa-garching.mpg.de/Virgo}}.

\printindex
\end{document}